\documentclass[twocolumn,amsmath,amssymb, aps,pra]{revtex4-1}

\pdfoutput=1

\usepackage{graphicx}
\usepackage[usenames]{color}
\usepackage{dcolumn}
\usepackage{bm}
\usepackage{hyperref}
\usepackage[mathlines]{lineno}

\hypersetup{
	colorlinks = true,
	urlcolor   = blue,
	linkcolor  = blue,
	citecolor  = blue
}

\DeclareMathOperator{\Tr}{Tr}
\newcommand*{\bra}[1]{\langle #1 |}
\newcommand*{\ket}[1]{| #1 \rangle}
\newcommand*{\argmax}{\mathop{\mathrm{argmax}}}

\begin{document}

\title{Experimental adaptive process tomography}

\author{I.\,A.\,Pogorelov$^1$}
	\email{Pogorelov5@yandex.ru}
\author{G.\,I.\,Struchalin$^1$}
\author{S.\,S.\,Straupe$^1$}
\author{I.\,V.\,Radchenko$^{1,2}$}
\author{K.\,S.\,Kravtsov$^{1,2}$}
\author{S.\,P.\,Kulik$^1$}
\affiliation{$^1$Faculty of Physics, M.\,V.\,Lomonosov Moscow State University, Moscow, Russia}
\affiliation{$^2$A.\,M.\,Prokhorov General Physics Institute RAS, Moscow, Russia}

\date{\today}

\begin{abstract}

Adaptive measurements were recently shown to significantly improve the performance of quantum state tomography. Utilizing information about the system for the on-line choice of optimal measurements allows to reach the ultimate bounds of precision for state reconstruction. In this article we generalize an adaptive Bayesian approach to the case of process tomography and experimentally show its superiority in the task of learning unknown quantum operations. Our experiments with photonic polarization qubits cover all types of single-qubit channels. We also discuss instrumental errors and the criteria for evaluation of the ultimate achievable precision in an experiment. It turns out, that adaptive tomography provides a lower noise floor in the presence of strong technical noise.

\end{abstract}

\pacs{03.65.Wj, 03.67.-a, 02.50.Ng, 42.50.Dv}

\maketitle

\section{Introduction}

Quantum information science commonly describes transformations of quantum states using a black-box approach~--- the details of the evolution are irrelevant and only the input-output relations are specified. This is natural in quantum communication, where a black box corresponds to a communication channel. Another natural situation where this point of view may be adopted is benchmarking and debugging quantum logic gates, which should be designed to produce specific outputs for given inputs. Mathematically this picture is described by the formalism of \emph{quantum processes}: given an input state $\rho$, the action of a quantum process is a completely-positive map $\rho\rightarrow \rho'=\mathcal{E}(\rho)$. The general task of \emph{quantum process tomography} is to reconstruct the (super)operator $\mathcal{E}$ form the experimental data provided by the outcomes of measurements on specific probe states \cite{Nielsen_JMO97,Zoller_PRL97,NielsenChuang}. The most straightforward way to achieve this task, which we adopt in this work, is to perform state tomography on the transformed probe states and derive the description of the process from this data. Adopting the terminology of \cite{Lidar_PRA08}, from now on we will focus on this \emph{standard quantum process tomography}.   

Since standard process tomography essentially utilizes state tomography as a subroutine, it seems natural, that it should benefit from the advanced methods of quantum state reconstruction. One of the recent major achievements in the field of quantum state tomography is the experimental implementation of adaptive measurement strategies \cite{Kravtsov_PRA13,Steinberg_PRL13,Kulik_PRA16,Ferrie_PRL16,Guo_NPJQI16} which allowed to qualitatively improve the precision of reconstruction. Adaptive strategies optimize subsequent measurements according to the current information about the state. It turns out, that such an optimization allows for a quadratic improvement in estimation quality over standard state tomography protocols for the same number of measurements \cite{Kravtsov_PRA13,Steinberg_PRL13}. Although the concept of self-learning measurements was known for a while \cite{Freyberger_PRA00,Wunderlich_PRA02}, only recent advances in computational methods allowed to create fast algorithms for online optimization in the course of experiment \cite{Houlsby_PRA12,Granade_NJP12,Ferrie_PRL14,Hen_NJP15}. 

In this Article we further develop the self-learning approach to make it applicable to quantum process tomography. Although quantum process tomography is mathematically related to state tomography via the Choi-Jamio\l kowski isomorphism, not any adaptive strategy will straightforwardly provide advantage, when applied to processes. As we explain below, the experimentally realizable strategies for process tomography correspond to a specific subclass of factorized measurements, so the adaptive strategy should show superior performance even when the optimization is restricted to this specific subclass. Fortunately, the algorithm, developed in \cite{Houlsby_PRA12} exhibits exactly this behavior. We also discuss the reconstruction of trace-non-preserving processes, which are useful for the description of lossy channels and require some additional care in reconstruction. Special attention is paid to the treatment of instrumental errors and quantification of maximal achievable precision in a real experiment. 

The Article is organized as follows: in Section~\ref{sec:theory} we review and provide all necessary information about the description of quantum processes, Bayesian approach to quantum tomography, and present the adaptive strategy; Section~\ref{sec:numerics} presents the results of our numerical simulations and discusses the influence of technical noise in measurements; experimental results are reported in Section~\ref{sec:experiment}; Section~\ref{sec:conclusion} concludes the paper. Technical details of the algorithm and experimental implementation are provided in Appendices.  

\section{Bayesian process tomography \label{sec:theory}}
\subsection{Introduction to quantum processes}
Quantum operations, also known as quantum processes or channels, are used to describe the evolution of quantum systems~\cite{NielsenChuang}. One of the approaches to the description of quantum processes is the \emph{operator-sum representation}. The action of a quantum operation $\mathcal{E}$ on a state $\rho$ can be represented as follows~\cite{Choi1975}:
\begin{equation}
	 \rho'=\mathcal{E}(\rho) = \sum\limits_{k=1}^{K} E_k \rho E_k^\dagger. \label{KrausForm}
\end{equation}
The number of terms~$K$ in the sum can be arbitrarily large, but it is always possible to limit their number by $d^2$ redefining the operators~$E_k$, $1 \le K \le d^2$, where~$d$ is the dimension of the state space of the system on which the process acts. That is why we will assume $K = d^2$ throughout the paper, unless the opposite is specified explicitly. The quantity $\Tr \mathcal{E}(\rho)$ defines loss in the channel. The \emph{operation elements} $\{E_k\}$ satisfy the requirement of trace-preservation $\sum_k E_k^\dagger E_k = I$ for trace-preserving processes or trace-non-increase $\sum_k E_k^\dagger E_k \leq I$ for processes with loss, in order to guarantee that $\Tr \mathcal{E}(\rho) \leq 1$. 

A lossless channel is a completely positive trace-preserving (CPTP) map from the operators~$\rho$ acting on the Hilbert space~$\mathcal H$ to the operators~$\rho'$ acting on the Hilbert space~$\mathcal H'$. For the sake of simplicity we will assume that $\dim \mathcal H = \dim \mathcal H' = d$, therefore the two spaces are isomorphic: $\mathcal H \cong \mathcal H'$. According to the Stinespring dilation theorem~\cite{Stinespring_PAMS55} the action of the CPTP map~$\mathcal E$ is equivalent to applying some unitary~$U$ to the extended system $\mathcal S \otimes \mathcal H$, followed by partial tracing over the auxiliary subsystem $\mathcal S$:
\begin{equation}
\rho' = \mathcal E(\rho) = \Tr_{\mathcal S} [U (\ket{0}\bra{0} \otimes \rho) U^\dagger], \label{eq:StinespringDilation}
\end{equation}
where~$\ket{0}$ is an arbitrary vector in the Hilbert space of $\mathcal S$. It is sufficient to choose $\dim \mathcal S = d^2$ to guarantee that the representation~(\ref{eq:StinespringDilation}) exists for any CPTP map.

In the standard computational basis one can easily compose the unitary matrix~$U$ from the elements~$\{E_k\}$ as the following block matrix of $d^3 \times d^3$ size:
\begin{equation}
U=\begin{pmatrix}
(E_1) & \ldots & \ldots & \ldots\\
(E_2) & \ldots & \ldots & \ldots\\
(E_3) & \ldots & \ldots & \ldots\\
\vdots & \vdots & \vdots & \vdots &
\end{pmatrix}
\label{BlockMatrixForm}.
\end{equation}
The first ``block column'' of the matrix $U$ determines the evolution of the principal system~$\mathcal H$, while the rest of this unitary matrix can be arbitrary.  

\paragraph*{$\chi$-matrix representation.} Let $\{\tilde{E}_k\}$ form a basis for the set of operators $\{E_k\}$, so that $E_k = \sum_{m=1}^{d^2} e_{km} \tilde{E}_m$, where $e_{km}$ are complex numbers. The equation (\ref{KrausForm}) turns into
\begin{equation}
	\mathcal{E}(\rho)=\sum\limits_{m,n=1}^{d^2}\tilde{E}_m \rho \tilde{E}_n^\dagger \chi_{mn}, \label{ChiMatrixForm}
\end{equation}
where the coefficients $\chi_{mn}=\sum_{k=1}^{d^2}e_{km}e_{kn}^*$ are the matrix elements of some Hermitian positive semidefinite matrix~$\chi$ by construction. The so called \emph{$\chi$-matrix representation}~(\ref{ChiMatrixForm}) completely determines the process $\mathcal{E}$, as well as the operator-sum representation~(\ref{KrausForm}). The $\chi$-matrix is often used for tomography purposes because it is more convenient to work with $d^4$ numbers~$\chi_{mn}$ rather than with $d^2$ matrices $E_k$ of $d \times d$ size.

The rank of the $\chi$-matrix is equal to the number of terms in~(\ref{KrausForm}). It is easy to see that a rank-1 trace-preserving process is a unitary process. We can define a \emph{purity} of the $\chi$-matrix $p = \Tr (\chi^2) / (\Tr \chi)^2$, analogously to the case of density matrices, to monitor a ``degree of unitarity''. For unitary processes $p=1$, while $p<1$ for non-unitary channels.

Another useful quantity is $\Tr \chi$, which is connected to the average loss in the channel~$\mathcal E$. Suppose a state~$\rho$ passes through the channel, then $\Tr \mathcal E(\rho)$ is the transmittance for the given state $\rho$. Using the representation~(\ref{ChiMatrixForm}) and integrating over the input states, the following expression for the average transmittance can be obtained:
\begin{equation}
\int \Tr \sum_{m,n} \tilde{E}_m \rho \tilde{E}_n^\dagger \chi_{mn} d\rho = \sum_{n} \frac{\chi_{nn}}{d} = \frac1d\Tr \chi,
\end{equation}
where we take into account that $\Tr \tilde{E}_m \tilde{E}_n^\dagger = \delta_{mn}$, and assume that the mean of $\rho$ with respect to the integration measure $d\rho$ is $\int \rho d \rho = 1/d$. For example, this assumption is valid for unitary invariant (Haar) measures $d\rho$, which are usually treated as ``uniform'' or uninformative~\cite{Granade_NJP16}. Therefore, the average loss in the channel is $1-\Tr\chi/d$.

\paragraph*{Choi-Jamio\l kowski isomorphism.} The $\chi$-matrix representation is closely related to the \emph{Choi-Jamio\l kowski isomorphism} \cite{Choi1975,Jamiokowski1972} between trace-preserving quantum operations $\mathcal{E}$ and density matrices $\rho_\mathcal{E}$ of the specific form in the extended space of dimension $d^2$:
\begin{equation}
	\rho_\mathcal{E}= [\mathcal{E}\otimes\mathcal{I}](\ket{\Psi}\bra{\Psi}), \label{CJIso}
\end{equation}
where $\ket{\Psi}=\sum_{j=1}^{d}\ket{j}\otimes\ket{j}/\sqrt{d}$ is a maximally entangled state and $\mathcal{I}$ is the identity operation acting trivially on the second subsystem. Different choices of the basis elements $\tilde E_m$ in~(\ref{ChiMatrixForm}) are possible. A convenient option is to select $\tilde{E}_{m = ld+l'} = \ket{l}\bra{l'}$, here $l, l' = 1, \dots, d$. In this basis the $\chi$-matrix of the process $\mathcal{E}$ is equal to its Choi-Jamio\l kowski state multiplied by $d$: $\chi = d \times \rho_\mathcal{E}$.

Therefore one can reduce process tomography to state tomography by preparing the bipartite state~$\ket{\Psi}$ in the extended system and passing one of its components through the channel $\mathcal E$. Tomography of the resulting state ~$\rho_\mathcal{E}$ reveals the $\chi$-matrix of the process. This procedure is called \emph{ancilla-assisted process tomography} (AAPT)~\cite{D'Ariano_PRL01,Leung_JMP03,Altepeter2003}.

\paragraph*{Process metrics.} Choi-Jamio\l kowski isomorphism offers an easy way to choose a metric to compare two quantum processes~\cite{Gilchrist2005}. A metric~$\Delta$ between two processes~$\mathcal E$ and $\mathcal F$ can be defined as some distance between the corresponding Choi-Jamio\l kowski states~$\rho_{\mathcal E}$ and $\rho_{\mathcal F}$:
\begin{equation}
	\Delta(\mathcal{E}, \mathcal{F}) = d(\rho_\mathcal{E},\rho_\mathcal{F}). \label{eq:ProcessMetricRho}
\end{equation}
This approach, though being powerful, is appropriate only for trace-preserving processes~--- the case when Choi-Jamio\l kowski isomorphism is applicable. For example, suppose an experimenter is interested in a polarization transformation in some channel and performs ancilla-assisted process tomography. If the channel has polarization independent loss (e.g. a neutral density filter in optics) then the experimenter will find that Choi-Jamio\l kowski states are the same for different values of loss. The distance~(\ref{eq:ProcessMetricRho}) will be equal to zero, but obviously the channels are different. A $\chi$-matrix gives a full description of the process and does not suffer from this deficiency. Moreover  $\chi$-matrices share the main properties with density matrices: a $\chi$-matrix is a Hermitian positive semidefinite matrix (however, $\Tr \chi \le d$ with equality holding for trace-preserving processes). Consequently, most of the widely used state metrics remain valid, if one substitutes $\chi$-matrices instead of density matrices.

In this paper we define a distance~$\Delta$ between two processes~$\mathcal E$, $\mathcal F$ as a Bures distance between the corresponding $\chi$-matrices:
\begin{equation}
\Delta(\mathcal{E}, \mathcal{F}) = d_B(\chi_\mathcal{E},\chi_\mathcal{F}), \label{eq:ProcessMetricChi}
\end{equation}
where the Bures distance is introduced as follows~\cite{Zyczkovsky_book_2006}:
\begin{equation}
d^2_B(A, B) =\Tr A + \Tr B - 2\Tr\sqrt{\sqrt{A}B\sqrt{A}}. \label{eq:BuresMetric}
\end{equation}
We note the appearance of $\Tr A + \Tr B \ne 2$, which contrasts the familiar definition for the quantum states with unit trace.

For trace-preserving processes both approaches~(\ref{eq:ProcessMetricRho}) and~(\ref{eq:ProcessMetricChi}) are suitable and there is a simple relation between them:
\begin{equation}
d_B(\rho_\mathcal{E},\rho_\mathcal{F}) = \frac{d_B(\chi_\mathcal{E},\chi_\mathcal{F})}{d},
\end{equation}
here~$d$ is the dimension of the principal system space.

\subsection{State and process tomography \label{trace-preserving}}
Let us first consider quantum state tomography. We will describe measurements using \emph{positive operator-valued measures} $\mathcal{M}_\alpha$ (POVMs), where the generic parameter $\alpha$ denotes the configuration of the experimental setup corresponding to the specific POVM. $\mathcal{M}_\alpha=\{M_{\alpha \gamma}\}$, where POVM elements $M_{\alpha \gamma}$ correspond to the particular measurement outcome $\gamma$, e.g. a count of a detector. The probability $\mathbb{P}$ of obtaining the result $\gamma$ having the system in the state $\rho$ and the experimental apparatus in the configuration $\alpha$ is given by the Born's rule:
\begin{equation}
	\mathbb{P}(\gamma|\rho,\alpha)= \Tr (M_{\alpha \gamma} \rho). \label{BornState}
\end{equation}
These probabilities can be estimated experimentally and the unknown state $\rho$ can be recovered after data analysis.

If one wants to recover an unknown process $\mathcal{E}$, he is allowed to vary the initial state $\rho_{\alpha}$ which the process acts on, in addition to varying the measurements $\mathcal{M}_\alpha$. It this case, utilizing the $\chi$-matrix representation (\ref{ChiMatrixForm}), we obtain
\begin{multline}
	\mathbb{P}(\gamma|\chi,\alpha)= \Tr (M_{\alpha \gamma} \mathcal{E}(\rho_\alpha)) =\\
	\Tr \Bigl(\sum\limits_{m,n=1}^{d^2}M_{\alpha \gamma} \tilde{E}_m \rho_\alpha \tilde{E}_n^\dagger \chi_{mn} \Bigr) =
	\Tr (M^\chi_{\alpha \gamma} \chi), \label{BornProc}
\end{multline}
where the matrix elements $(M_{\alpha \gamma}^\chi)_{nm} = \Tr (M_{\alpha \gamma} \tilde{E}_m \rho_\alpha \tilde{E}_n^\dagger)$. Selecting the basis $\tilde{E}_{l+dl'} = \ket{l} \bra{l'}$, one can obtain a simple relation: $M^\chi_{\alpha \gamma} = M_{\alpha \gamma} \otimes \rho^*_\alpha$, where $\rho^*_\alpha$ denotes a complex conjugate of $\rho_\alpha$. Therefore, the measurement operators $M^\chi_{\alpha \gamma}$ are always factorized in this sense.

The equations (\ref{BornState}) and (\ref{BornProc}) establish an explicit analogy between state and process tomography~\footnote{However the set $\{M^\chi_{\alpha \gamma}\}$ does not form the decomposition of unity, $\sum_{\gamma} M^\chi_{\alpha \gamma} \ne I$, unlike the case of state tomography, where it is usually assumed that $\sum_{\gamma} M_{\alpha \gamma} = I$}. The differences are the size of the matrix recovered: $d^2 \times d^2$ for a process $\chi$-matrix and $d \times d$ for a state density matrix~$\rho$, and the restriction to factorized measurements, described above.
 
\subsection{Bayesian approach}
After the measurements are performed, one should process the data obtained. An \emph{estimator} must be received as the result of this processing. We consider the Bayesian approach \cite{BlumeKohout_NJP10,Granade_NJP16} for estimation of the unknown matrix. The Bayesian approach works with the probability distribution over the space of process matrices $p(\chi|\mathcal{D})$, where $\mathcal{D}$ denotes the set of outcomes $\{\gamma_n\}$. This probability can be calculated via Bayes' rule:
\begin{equation}
	p(\chi|\mathcal{D}) \propto \mathcal{L}(\chi;\mathcal{D}) p(\chi|\varnothing). \label{Posterior}
\end{equation}
Here $\mathcal{L}(\chi;\mathcal{D}) = \prod_{n} \mathbb{P}(\gamma_n|\chi,\alpha_n)$ is a \emph{likelihood function} and $p(\chi|\varnothing)$ is a \emph{prior distribution} which reflects our preliminary knowledge about the system of interest. One can use \emph{Bayesian mean estimate} (BME) to recover an unknown matrix
\begin{equation}
	\hat{\chi}=\int \chi p(\chi|\mathcal{D})d\chi. \label{BayesAve}
\end{equation}
The uncertainty of such an estimator can be assessed via the \emph{distribution size} in a particular metric (\ref{eq:BuresMetric})
\begin{equation}
	\overline{d}^2_B = \int d_B^2(\chi,\hat{\chi}) p(\chi | \mathcal D) d\chi. \label{BayesDistrSize}
\end{equation}


\subsection{Adaptivity}

Another important point in tomography is a measurement sequence. We try to construct the measurement sequence in the most effective way to guarantee a better and faster reconstruction of the unknown process. In the Bayesian approach the posterior distribution allows one to use Shannon entropy decrease criterion \cite{Houlsby_PRA12} to choose the next measurement in an optimal way. The following relation can be used to find the setup configuration $\alpha_\text{next}$ corresponding to the best next measurement:
\begin{equation}
	\alpha_\text{next}=\argmax_{\alpha}\{ \mathbb{H}[\mathbb{P}(\gamma|\alpha, \mathcal D)]-\mathbb{E}_{p(\chi|\mathcal{D})}\mathbb{H}[\mathbb{P}(\gamma|\chi,\alpha)]\}, \label{eq:AdaptiveEntropy}
\end{equation}
where $\mathbb{H}$ is the Shannon entropy, $\mathbb{P}(\gamma|\alpha, \mathcal D) = \int \mathbb{P}(\gamma| \chi, \alpha) p(\chi | \mathcal D) d\rho$, and $\mathbb{E}_{p(\chi|\mathcal{D})}$ denotes the average over $p(\chi|\mathcal{D})$. Such a measurement strategy depends on the data collected through previous measurements, so it is \emph{adaptive}. We will compare the adaptive strategy (A) with the sequence of randomly chosen measurements (R).

\section{Numerical simulations \label{sec:numerics}}

\subsection{Adaptivity benefit}

\begin{figure}
	\centering
	\includegraphics[width=1.0\linewidth]{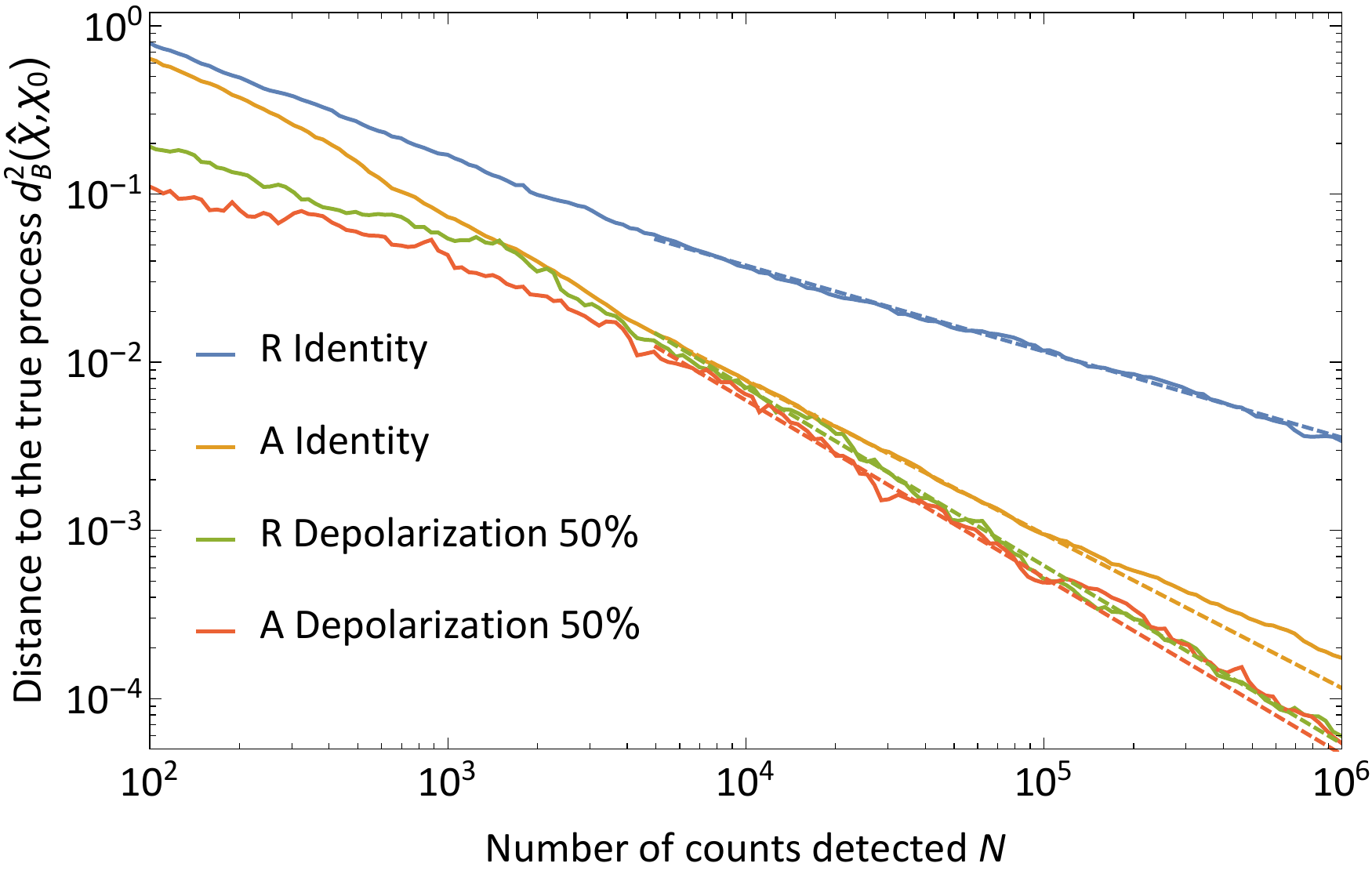}
	\caption{Evolution of the distance $d^2_B(\hat{\chi},\chi_0)$ from the estimator~$\hat \chi$ to the true process $\chi_0$ with the number of events $N$ registered. Two processes are studied: an identity process and a depolarizing channel with 50\% depolarization. The results are averaged over 10 runs. Here and on the consecutive plots ``R'' stands for the random and ``A'' stands for the adaptive measurement strategy.}
	\label{fig:sim_adaptivity_profit_1}
\end{figure}

Before reporting the experimental results let us present numerical simulations. First of all, the performance of tomography for different processes was studied. We quantify the performance by the Bures distance (\ref{eq:BuresMetric}) of the estimate, i.e. the Bayesian average (\ref{BayesAve}), to the true process. Typical evolution of the distance to the true process $d_B^2(N)$ on the number of counts detected~$N$ is shown in Fig.~\ref{fig:sim_adaptivity_profit_1}. One can see that adaptive measurements have an advantage over random measurements for the identity process~--- the process having no effect on the polarization and leaving the initial state undisturbed. The situation is the same for other unitary transformations, e.g. for a wave plate. However, the advantage disappears for non-unitary processes like a depolarization channel.

\begin{table}[b]
	\caption{\label{tab:table1}
		Approximation of the distribution size dependence~$\overline{d}^2_B(N)$ of the number of photons detected~$N$, obtained in simulations, with $C N^\alpha$ model. The processes are considered as trace-preserving ones. 
	}
	\begin{ruledtabular}
		\begin{tabular}{lccc}
			\textrm{Meas. strategy \& process}&
			\textrm{$\alpha$}&
			\textrm{$C$}\\
			\colrule
			R Identity & $-0.5119 \pm 0.0015 $ & $1.436 \pm 0.015 $  \rule{0pt}{11pt}\\ 
			A Identity & $-0.9158 \pm 0.0016 $ & $3.585 \pm 0.015 $ \\
			R Depolarization 50\% & $-1.060 \pm 0.005 $ & $4.81 \pm 0.05 $ \\
			A Depolarization 50\% & $-1.053 \pm 0.004 $ & $4.58 \pm 0.04 $ \\
		\end{tabular}
	\end{ruledtabular}
\end{table}

Table~\ref{tab:table1} shows power law fits $C N^\alpha$ of the dependence $d_B^2(N)$. The \emph{convergence rate}~$\alpha \approx -1$ for the adaptive protocol, regardless of the true process. For the random measurement sequence the identity process and other rank-1 (i.e. unitary) channels are hard to estimate ($\alpha \approx -0.5$). These results are analogous to the case of quantum state tomography \cite{Houlsby_PRA12,Kulik_PRA16} where the convergence rate $\alpha \approx -0.5$ was also observed for random measurements of pure (i.e. rank-1) states, and there was no adaptivity benefit for mixed states ($\alpha \approx -1$ for both adaptive and random protocols).

\subsection{Influence of noise \label{Noise}}
\begin{figure}
	\centering
	\includegraphics[width=1.0\linewidth]{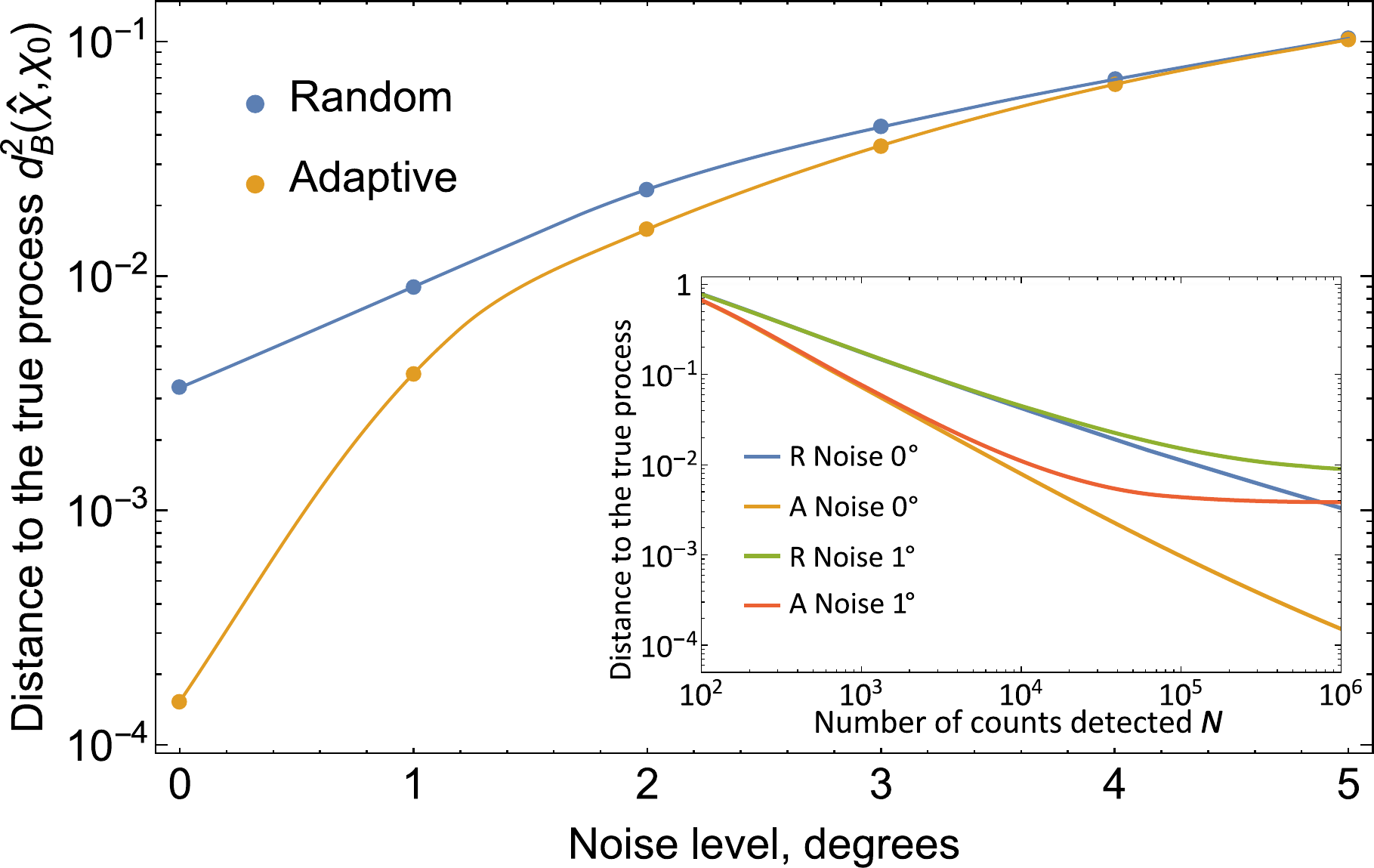}
	\caption{Dependence of the distance to the true process $d^2_B(\hat{\chi},\chi_0)$, taken after $N=10^6$ events were detected, with the noise level in wave plates positions. The adaptive strategy (yellow dots) has advantage over random measurements (blue dots). Solid lines are guides to the eye. Inset: the evolution of the distance to the true process $d^2_B(\hat{\chi},\chi_0)$ with the number of events registered $N$ for noiseless measurements (Noise 0$^\circ$) and for $\phi_0 = 1^\circ$ noise level (Noise 1$^\circ$). All results are averaged over 1000 tomography runs.}
	\label{fig:sim_noise_variation_10000s_5}
\end{figure}

\begin{figure}
\centering
\includegraphics[width=1.0\linewidth]{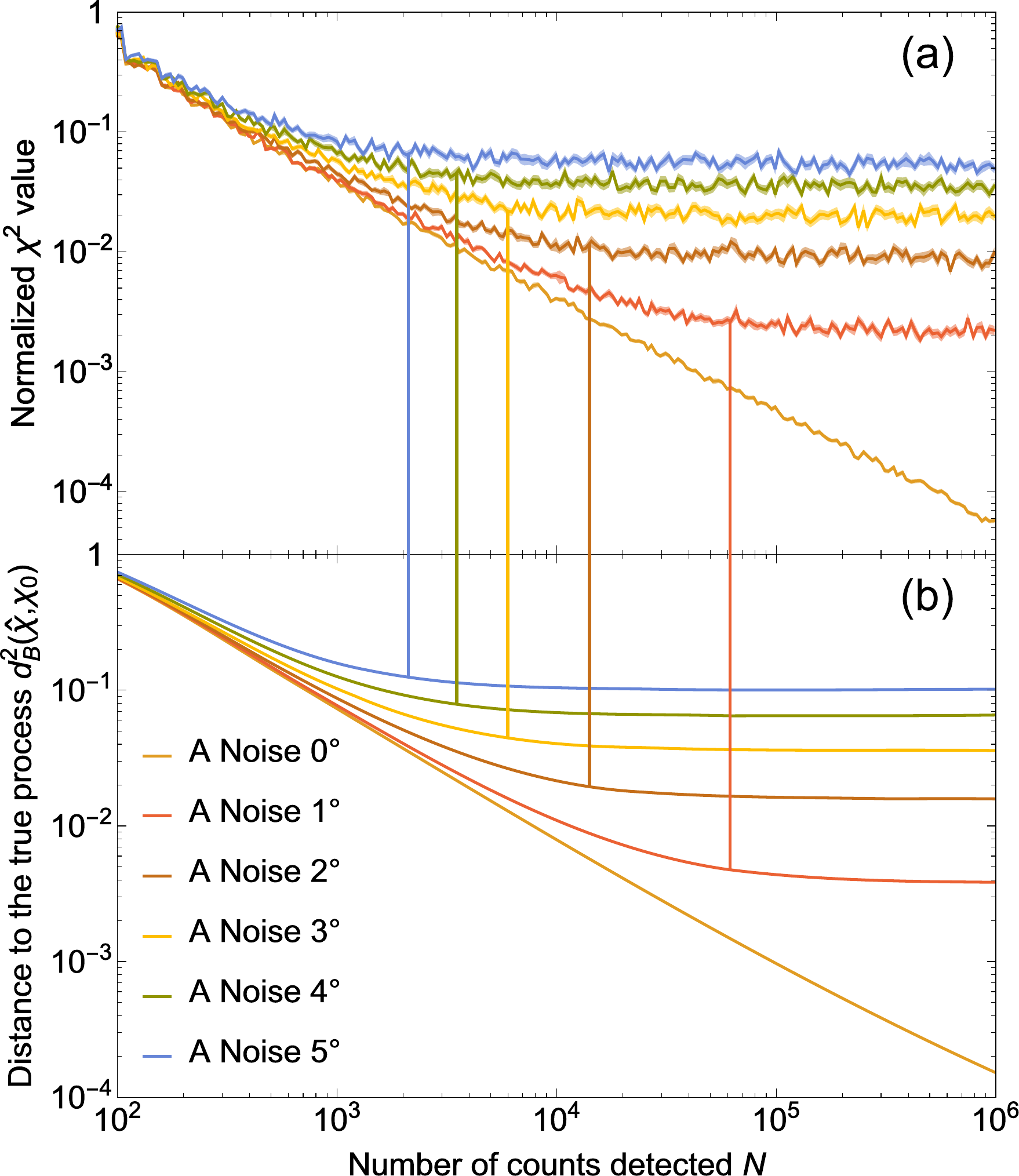}
\caption{Comparison between (a) the dependence of the normalized chi-squared test statistic~$\chi^2/b$ on the number of events registered~$N$ and (b) the distance to the true process~$d_B^2(\hat\chi, \chi_0)$. The results shown are the numerical simulations for the adaptive measurements with various noise levels. The true process is the identity channel. The vertical lines show the values of~$N$ for which the double logarithmic derivative $\frac{d\ln d_B^2(\hat\chi, \chi_0)}{d \ln N} = -0.25$ for the corresponding noise levels. All results are averaged over 1000 runs.}
\label{fig:comparison}
\end{figure}

Any experiment usually suffers from some imperfections. In our experiments with polarization qubits we use wave plates equipped with rotation motors. Hence we simulate errors in wave plates angle settings. The error is modeled by a uniform distribution, i.e. after setting the wave plate, its position is shifted for a random angle $\Delta \phi\in[-\phi_0,\phi_0]$. This causes tomography to stop converging at a certain moment and reach its limit of accuracy \cite{Genovese11}. An example of such a behavior is represented in the inset of Fig.~\ref{fig:sim_noise_variation_10000s_5}, where noiseless tomography ($\phi_0 = 0^\circ$) is compared with noisy measurements ($\phi_0 = 1^\circ$). Here another advantage of adaptivity is revealed: adaptive measurements have higher ultimate accuracy level than random ones. This can be explained by the specific features of adaptive measurements \cite{Kulik_PRA16}.

Fig.~\ref{fig:sim_noise_variation_10000s_5} compares the distance of the current estimate $\hat{\chi}$ to the true process~$d_B^2(\hat \chi, \chi_0)$ after the fixed number~$N$ of registered events for different noise levels~$\phi_0$. The identity channel is chosen as the true process. We select $N = 10^6$ because it was found sufficient to achieve the ultimate accuracy level for noise amplitudes $\phi_0 \gtrsim 1^\circ$. The advantage of the adaptive protocol is more evident for low noise amplitudes and vanishes for $\phi_0 \gtrsim 4^\circ$.

\paragraph*{Stopping criterion.\label{RJD}}
Generally, it is impossible to measure the distance to the true process in the experiment. As it can be seen in the inset of Fig.~\ref{fig:sim_noise_variation_10000s_5}, for the adaptive protocol there is no sense to do any measurements after $10^5$ events have already been registered, because the noise limits the accuracy of the result. Therefore, some stopping criterion is required to recognize the moment when further measurements will provide no more information. Moreover, one should be able to apply this criterion without any knowledge of the true process, as in the real experiment. To attain this goal we utilized a well-known chi-squared test statistic~$\chi^2$, which was proven to indicate the consistency of the current estimate~$\hat \chi$ with the data observed~\cite{Mogilevtsev_PRL13}:
\begin{equation}
\chi^2 = \sum_{\gamma} \frac{(n_\gamma - b \hat p_\gamma)^2}{b \hat p_\gamma}, \label{eq:ChiSquared}
\end{equation}
where $n_\gamma$ is the number of events when the outcome~$\gamma$ was detected, $b = \sum_\gamma n_\gamma$ is the total number of events for a particular measurement configuration~$\alpha$, and $\hat p_\gamma = \mathbb P(\gamma | \hat \chi, \alpha)$ is the expected probability of the outcome~$\gamma$.

In our case we have two possible outcomes with probabilities~$p_0 = p$ and~$p_1 = 1 - p$. Thus, $n_0$ is a binomially distributed random variable with the mean $b p_0$: $n_0 \sim \text{Bin}(b, p_0)$. It is easy to calculate the mean of the chi-squared statistic~$\langle \chi^2 \rangle$:
\begin{equation}
\langle \chi^2 \rangle = \frac{b(p - \hat p)^2 + p(1-p)}{\hat p (1- \hat p)}.
\end{equation}
In the absence of noise the estimator converges to the true process, so $\hat p = p$ and $\langle \chi^2 \rangle = 1$ in the limit of large~$N$. Obviously, due to errors in the real apparatus one cannot expect the perfect convergence and $\langle \chi^2 \rangle \ne 1$. The measurement \emph{block size}~$b$ should be large enough to reliably determine the difference of~$\langle \chi^2 \rangle$ from unity in the experiment. We used $b \propto N$, which is a reasonable trade-off between the benefit from adaptivity and the overall measurement time~\cite{Kulik_PRA16}. Given this block size schedule, the second term in the nominator of~(\ref{eq:ChiSquared}) can be omitted  for $N \to \infty$, obtaining $\langle \chi^2 \rangle \propto b$. The normalized  quantity~$\chi^2/b$ converges to some constant value, depending on the noise magnitude. For noiseless measurements this value is zero.

Therefore one can judge about the convergence of the tomographic protocol by monitoring the normalized chi-squared test statistic $\chi^2/b$. When it reaches a constant value, the distance to the true process~$d_B^2(\hat \chi, \chi_0)$ also does, and the measurements should be stopped. We verified this in numerical simulations. A comparison of the dependencies of $\chi^2/b$ and $d_B^2(\hat \chi, \chi_0)$ on the number of events registered~$N$ is depicted in Fig.~\ref{fig:comparison} for various noise levels. One can clearly see that the noise floor for both quantities is achieved for the same values of~$N$.


\section{Experiment \label{sec:experiment}}

\subsection{Setup}
\begin{figure}
\centering
\includegraphics[width=1.0\linewidth]{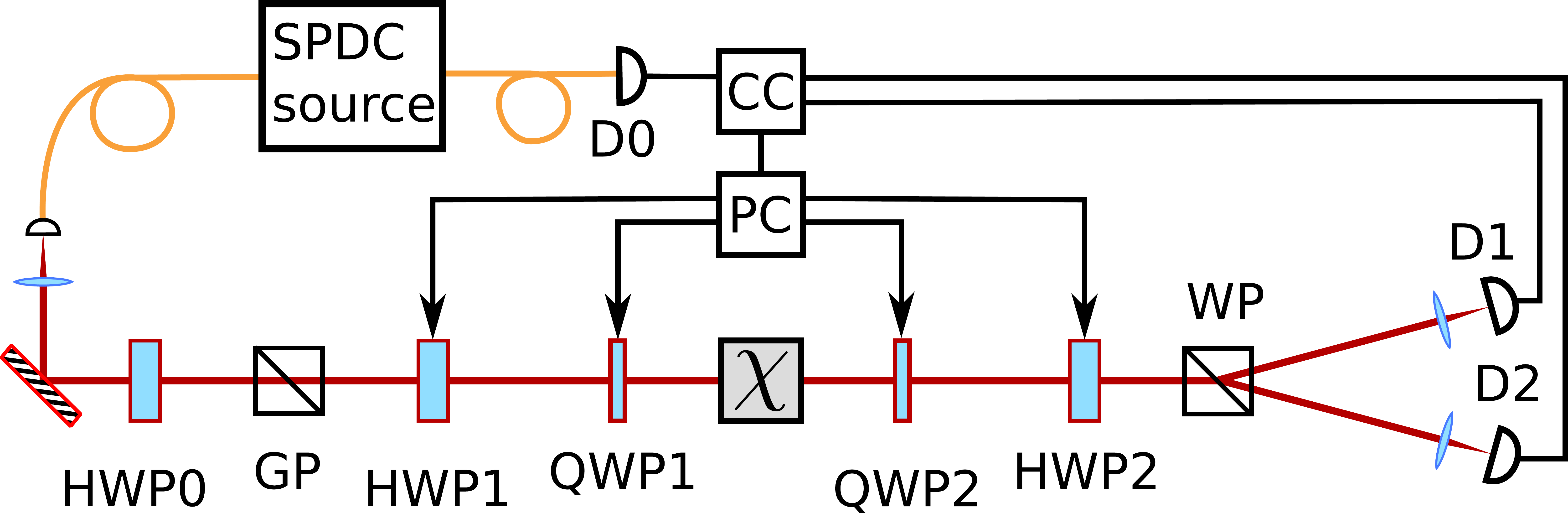}
\caption{Experimental setup. An SPDC source of photon pairs is used to prepare a heralded single photon state. Computer controlled wave plates allow us to prepare an arbitrary initial state and to make arbitrary projective measurements.}
\label{fig:setup_heralded}
\end{figure}

The tomography protocol is realized experimentally for quantum processes acting on single-photon polarization states. We used a heralded single photon source based on the spontaneous parametric down-conversion (SPDC) in a 25 mm long PPKTP crystal inside a Sagnac interferometer~\cite{Kim2006a}. The setup is organized as follows (Fig.~\ref{fig:setup_heralded}). A Glan prism GP, wave plates HWP1 (half-wave) and QWP1 (quarter-wave) are used to prepare the initial state. Wave plates QWP2, HWP2 and a Wollaston prism WP allow us to perform arbitrary projective measurements after the unknown process $\chi$ acts on the initial state. The first photon from the pair passes through the elements described and is coupled to the multimode fibers leading to the single-photon counting modules (SPCMs) D1-2. The second one is used as a trigger being detected by the SPCM D0. All wave plates are equipped with computer controlled motorized rotation stages to implement the active measurements. An unknown process $\chi$ may be represented by various optical elements, e.g. a polarizer, a wave plate, a multimode optical fiber, etc.

\subsection{Ultimate accuracy level}

\begin{figure}
	\centering
	\includegraphics[width=1.0\linewidth]{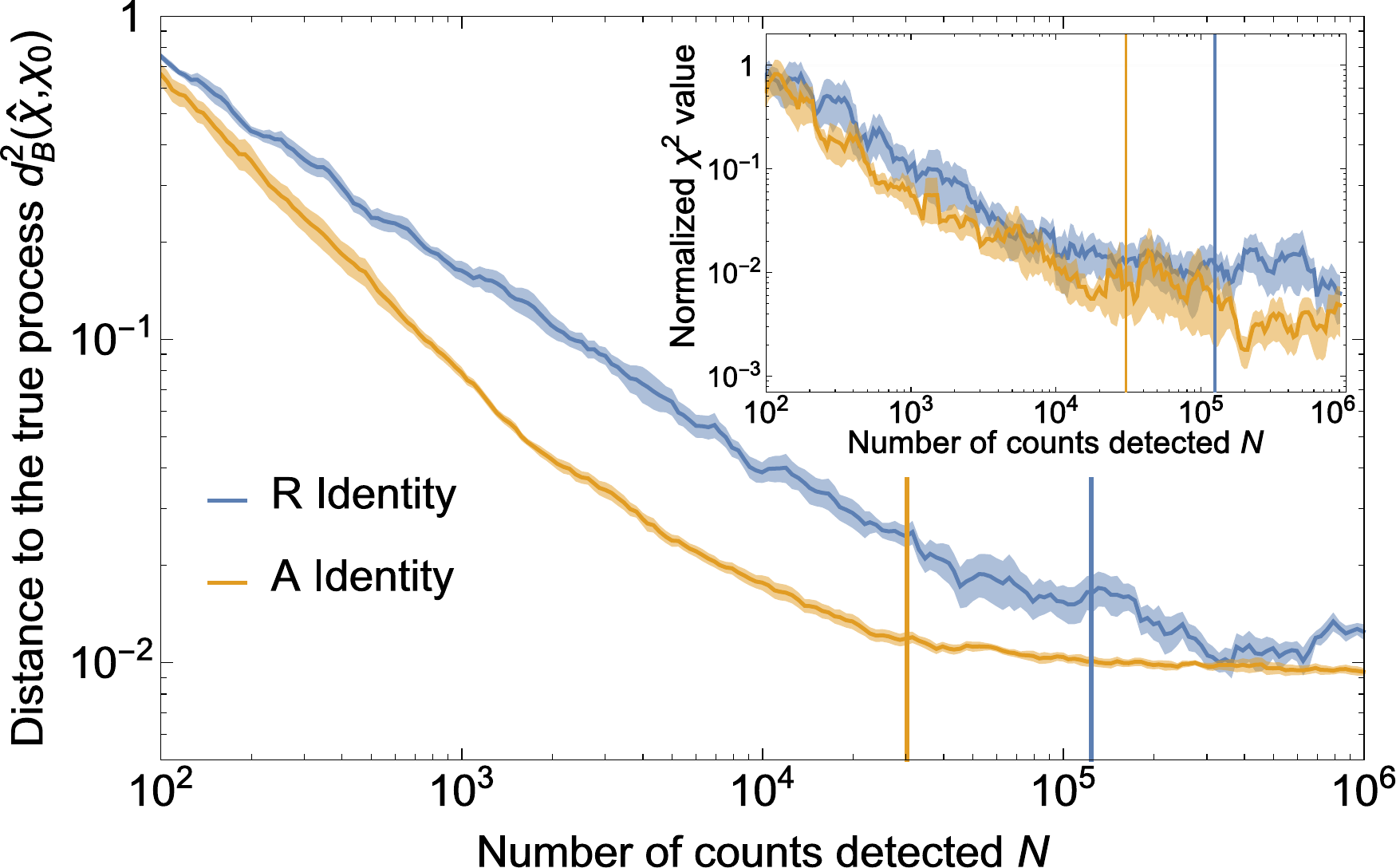}
	\caption{Experimental dependence of the Bures distance $d_B^2(\hat \chi, \chi_0)$ to the theoretical true process (an identity channel) with the number of photons detected~$N$ for random (R) and adaptive (A) measurements. Inset: the dependence of the normalized chi-squared test statistic $\chi^2/b$ on $N$. Vertical lines show values of~$N$ when the (smoothen) double logarithmic derivative $\frac{d\ln d_B^2(\hat\chi, \chi_0)}{d \ln N} = -0.25$ for the corresponding protocols. All results are averaged over 10 tomography runs. Additionally, $\chi^2/b$ was smoothened via a moving average along~$N$ axis, calculated for 5 successive points.}
	\label{fig:exp_dist_to_theor_1}
\end{figure}

As it was mentioned in Sec.~\ref{Noise}, no experimental setup is perfect. Some imperfections like wave plates retardance errors, Glan and Wollaston prism angle setting errors, detectors dark counts, etc. can result in a decrease of tomography accuracy. There is a moment when the measurements should be stopped, because tomography does not converge anymore due to the experimental imperfections. We studied the ultimate accuracy, which we can achieve in tomography for the case of an identity process. Such a process does not change the polarization, so experimentally it is realized by simply placing no elements in a process~$\chi$ placeholder in Fig.~\ref{fig:setup_heralded}. The identity process is a distinguished case in the experiment, because it can be prepared exactly, and the true $\chi$-matrix is known. So one can measure the distance between the theoretical $\chi$-matrix of an identity process and the Bayesian estimator obtained via tomography. The dependence of the distance on the number of photons detected is presented in Fig.~\ref{fig:exp_dist_to_theor_1}. The adaptive measurement strategy provides faster tomography convergence and higher ultimate accuracy level than the random measurements. The final Bures distance between the Bayesian mean estimator and the theoretical true process, acquired after $N = 10^6$ registered events is listed in Table~\ref{tab:expAccuracy} for both random and adaptive measurements. We attribute the achieved values of the noise floor mostly to the wave plates retardance errors and misalignments in their angular positioning.

\begin{table}[h]
	\caption{\label{tab:expAccuracy}
		Final Bures distance between the Bayesian mean estimate and the theoretical true process, obtained after $N = 10^6$ counts are detected, for the experimental tomography of an identity process. Fidelity is calculated for the corresponding Choi-Jamio\l kowski states.
	}
	\begin{ruledtabular}
		\begin{tabular}{lccc}
			\textrm{Meas. strategy}&
			\textrm{Bures distance, $d_B^2$}&
			\textrm{Fidelity, $F$}\\
			\hline		
			Random & $0.0125 \pm 0.0017 $ & $0.9938 \pm 0.0008 $	\rule{0pt}{11pt}\\ 
			Adaptive & $0.0094 \pm 0.0008 $ & $0.9953 \pm 0.0004 $ \\
		\end{tabular}
	\end{ruledtabular}
\end{table}

We also computed the normalized chi-squared test statistic~$\chi^2/b$, which is shown in the inset of Fig.~\ref{fig:exp_dist_to_theor_1}. It reaches an approximately constant value after $N \approx 3 \times 10^4$ detected photons, and according to the criterion proposed in Sec.~\ref{RJD} the measurements should be stopped at this point. The results are in a reasonable correspondence with the ones for the distance to the true process $d_B^2(\hat \chi, \chi_0)$. The disadvantage of this test statistic is that it fluctuates a lot from one tomography run to another and thus a large number of runs are required to achieve a smooth average.

\subsection{Adaptivity benefit}
\begin{figure}
\centering
\includegraphics[width=1.0\linewidth]{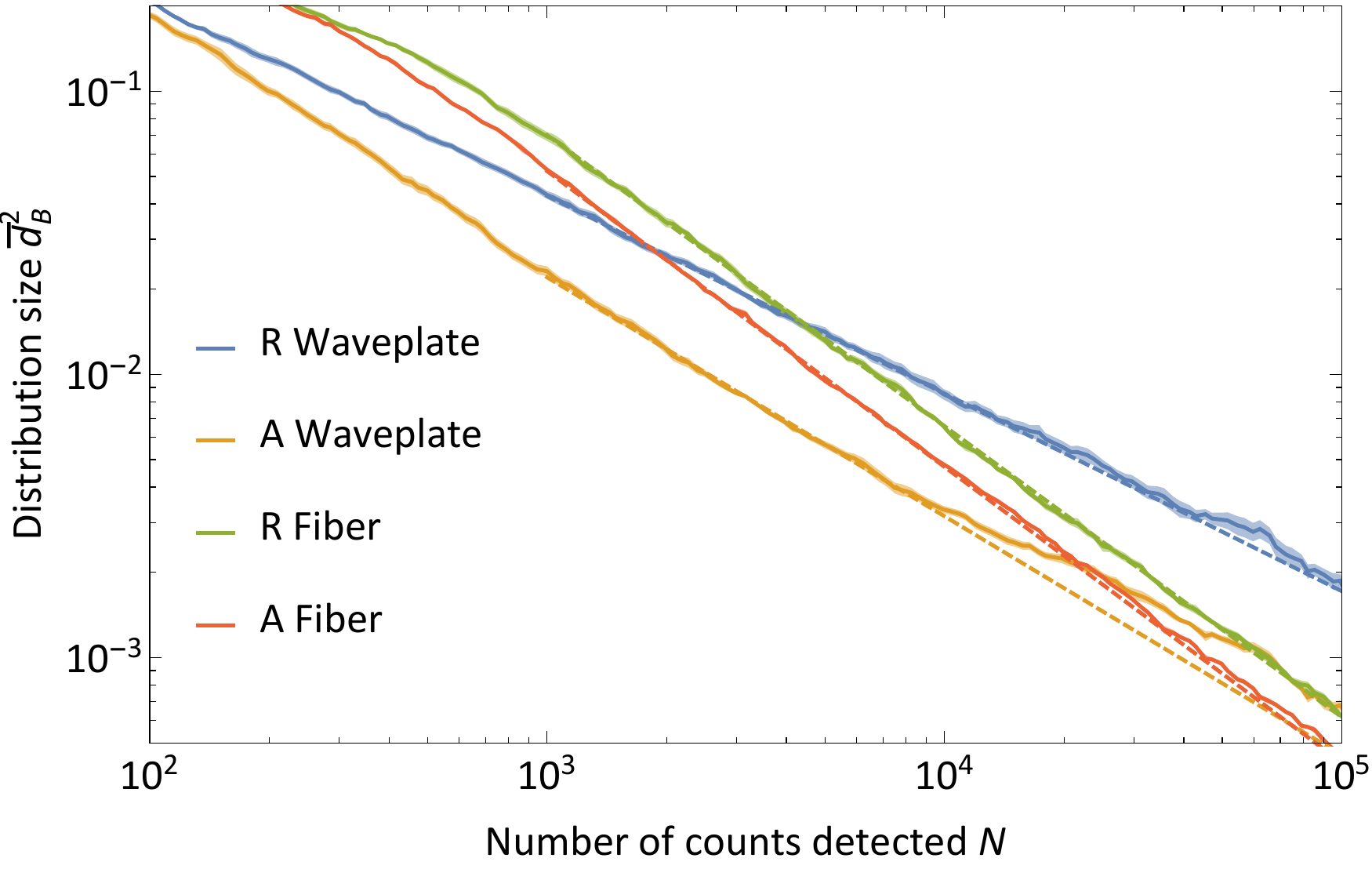}
\caption{Experimental evolution of the distribution size $\overline{d}^2_B$ with the number of photons detected $N$ for random (R) and adaptive (A) measurements. Two processes are studied: a unitary process (quarter wave plate) and a non-unitary depolarizing channel (multimode fiber). The results are averaged over 10 runs.}
\label{fig:exp_adaptivity_profit_2}
\end{figure}

In this section we experimentally study the difference between adaptive tomography and random measurements for unitary and non-unitary trace-preserving processes. A quartz quarter-wave plate is used as an example of a unitary process, while the multimode fiber represents a non-unitary process, because if the spatial modes are averaged out, it acts as a completely depolarizing channel (see appendix Sec.~\ref{sec:MMF}). Comparison of the distribution size dependence~$\overline{d}_B^2(N)$ on the number of detected photons~$N$ is depicted in Fig.~\ref{fig:exp_adaptivity_profit_2} for both processes and both measurement strategies. For a quantitative comparison we fit the dependence $\overline{d}_B^2(N)$ with a model of the form $C N^\alpha$. The results are listed in Table \ref{tab:expAdaProfit}. The adaptive measurement strategy has an advantage in convergence over the random measurements, and the advantage is much more significant for unitary processes.

\begin{table}[h]
	\caption{\label{tab:expAdaProfit}
		Fit of the experimental distribution size dependence $\overline{d}^2_B(N)$ on the number of detected photons $N$ with a $C N^\alpha$ model for random (R) and adaptive (A) measurements and two processes. Both processes are considered to be  trace-preserving.
	}
	\begin{ruledtabular}
		\begin{tabular}{lccc}
			\textrm{Meas. strategy \& process}&
			\textrm{$\alpha$}&
			\textrm{$C$}\\
			\hline		
			R Unitary (wave plate) & $-0.698 \pm 0.016 $ & $1.67 \pm 0.13 $ \rule{0pt}{11pt}\\ 
			A Unitary (wave plate) & $-0.844 \pm 0.016 $ & $2.01 \pm 0.13 $ \\
			R Depolarizing (fiber) & $-1.027 \pm 0.013 $ & $4.44 \pm 0.11 $ \\ 
			A Depolarizing (fiber) & $-1.044 \pm 0.006 $ & $4.26 \pm 0.05 $ \\
		\end{tabular}
	\end{ruledtabular}
	
\end{table}

\subsection{Trace non-preserving processes \label{lossy}}

\begin{figure}
\centering
\includegraphics[width=1.0\linewidth]{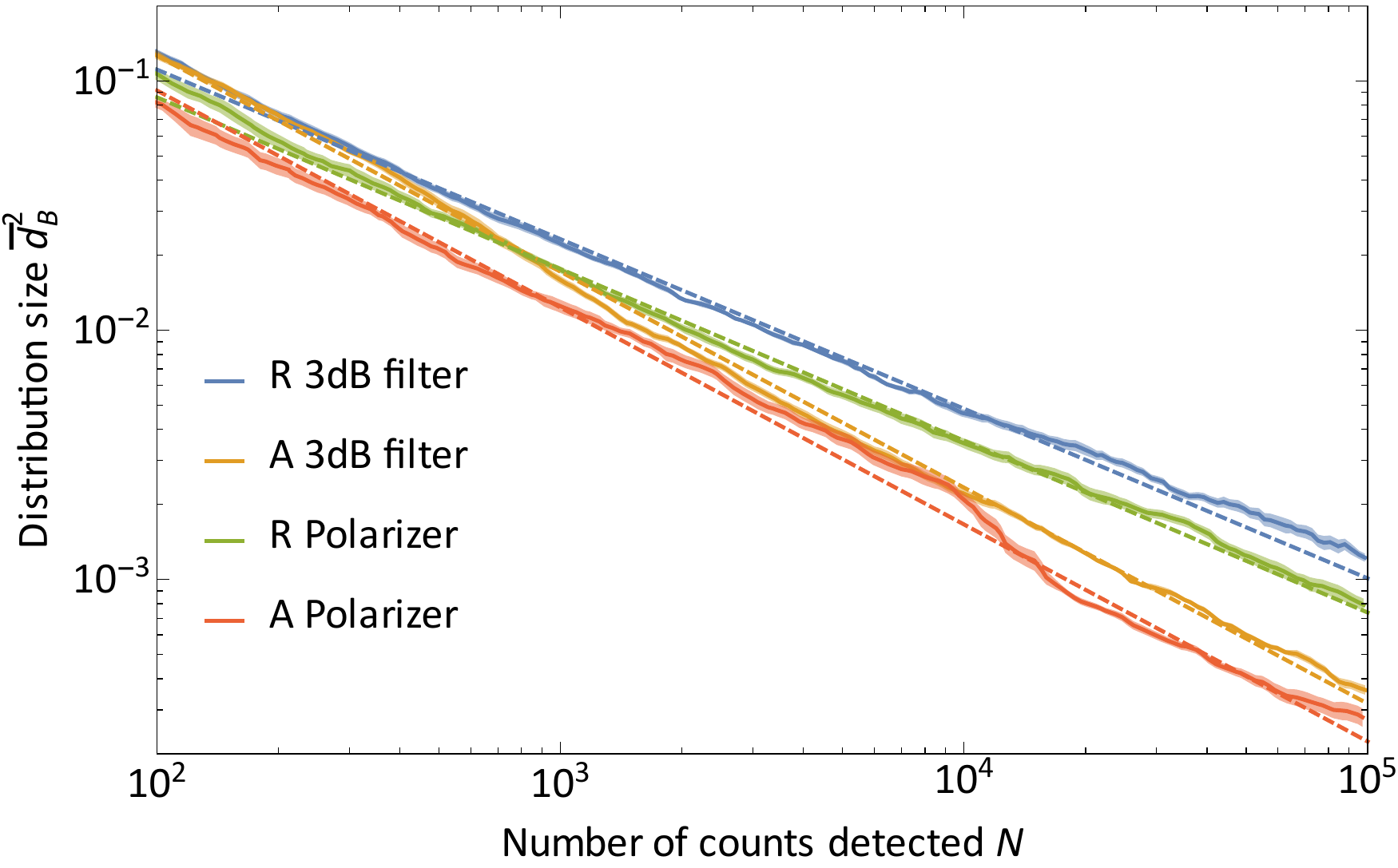}
\caption{Experimental evolution of the distribution size $\overline{d}^2_B$ with the number of photons detected $N$ for non-trace-preserving processes. Two processes are considered: a polarization independent lossy process (a 3~dB neutral density filter) and polarization-dependent loss (a linear polarizer). The results are averaged over 10 runs.}
\label{fig:exp_lossy_2}
\end{figure}

The tomography of trace non-preserving processes is a little bit more complicated than that of trace-preserving ones. Loss must be taken into account, so the expression~(\ref{BornProc}) for the probability $P(\gamma|\chi,\alpha)$ should be modified. The full intensity of the light incident on each detector, when no additional loss connected with the investigated process are present, should be known. In the experiment this can be done in the following way: the investigated process~$\chi$ is replaced with an identity channel and all photons are directed to one of the detectors (for example, corresponding to $\gamma=0$) by varying their polarization appropriately. The intensity~$I_{\gamma = 0}$ is measured and the procedure is repeated for the second detector, corresponding to $\gamma=1$.

If the measurement~$\alpha$ lasted for the time~$t$ then the probability to obtain a set of outcomes~$\{n_\gamma\}$ is:
\begin{equation}
	\mathbb{P}(\{n_\gamma\}|\chi,\alpha, t) \propto \prod_{\gamma=0}^{d-1} \mathbb{P}^{n_\gamma}(\gamma|\chi,\alpha) e^{-I_\gamma \mathbb{P}(\gamma|\chi,\alpha)t}. \label{Non-trace-preserving}
\end{equation} 
Here we supposed that the source of light has Poissonian statistics. Also, we omit the terms, which do not depend on~$\chi$, because they can be absorbed in the normalization of the posterior, hence the proportionality sign is used. Previously, in the case of trace-preserving processes, we had $\sum_{\gamma} I_\gamma \mathbb{P}(\gamma|\chi,\alpha)=\text{const}$, regardless of~$\chi$, so the probability (\ref{Non-trace-preserving}) was simply $\mathbb{P}(\{n_\gamma\}|\chi,\alpha, t) \propto \prod_\gamma \mathbb{P}^{n_\gamma}(\gamma|\chi,\alpha)$ and was independent of $t$.

The modified expression~(\ref{Non-trace-preserving}) should be substituted into~(\ref{eq:AdaptiveEntropy}) instead of~$\mathbb P(\gamma | \chi, \alpha)$ in order to calculate the next optimal measurement. However, in the case of trace-non-preserving processes the space of possible outcomes~$\{n_\gamma\}$ is infinitely large, therefore calculation of entropies $\mathbb H[ \mathbb{P}(\{n_\gamma\}|\chi,\alpha, t)]$ involves an infinite series summation. This is computationally intractable and we turned out with the following heuristics:
\begin{equation}
	\alpha_\text{next}=\argmax_{\alpha}\{
	 \mathbb{H}[\pi(\gamma|\hat \chi, \alpha)]-\mathbb{E}_{p(\chi|\mathcal{D})}\mathbb{H}[\pi(\gamma|\chi,\alpha)]
	\}, \label{eq:AdaptiveEntropyLoss}
\end{equation}
where $\pi(0 | \chi, \alpha) = \mathbb{P}(0 | \chi, \alpha)$ and $\pi(1 | \chi, \alpha) = 1- \mathbb{P}(0 | \chi, \alpha)$. We note that $\pi(1 | \chi, \alpha) \ne \mathbb{P}(1 | \chi, \alpha)$ because of the presence of loss. Utilizing this heuristics the complexity of calculations remains the same as for trace-preserving processes, which allows us to carry on-line adaptive measurements.

Experimental tomography, taking into account the details described above, was performed. We present the experimental results for three processes: identity, a 3\,dB neutral filter, as an example of polarization-independent loss, and a polarizer~--- a polarization-dependent lossy process. The processes were now considered as trace non-preserving ones. The results are shown in Fig.~\ref{fig:exp_lossy_2} and Table~\ref{tab:expLossy}. As one can see, all of these processes enjoy the advantage in convergence of adaptive tomography. We attribute this to the fact that all considered processes are rank-1 channels, although a polarizer and a 3\,dB neutral filter are not unitary ones. The main conclusion is that adaptive tomography provides better reconstruction accuracy for rank-1 channels regardless of the amount of loss.

\begin{table}[h]
	\caption{\label{tab:expLossy}%
		Approximation of the experimental distribution size dependence~$\overline{d}^2_B(N)$ of the number of photons detected~$N$ with a $C N^\alpha$ model. The processes are considered as trace non-preserving ones.
	}
	\begin{ruledtabular}
		\begin{tabular}{lccc}
			\textrm{Meas. strategy \& process}&
			\textrm{$\alpha$}&
			\textrm{$C$}\\
			\hline		
			R Identity\rule{0pt}{11pt} & $-0.630 \pm 0.003 $ & $1.21 \pm 0.03 $ \\ 
			A Identity & $-0.764 \pm 0.003 $ & $1.59 \pm 0.03 $ \\
			R 3\,dB neutral filter & $-0.680 \pm 0.003 $ & $0.94 \pm 0.02 $ \\ 
			A 3\,dB neutral filter & $-0.866 \pm 0.002 $ & $1.92 \pm 0.02 $ \\
			R Polarizer & $-0.689 \pm 0.004 $ & $0.72 \pm 0.03 $ \\ 
			A Polarizer & $-0.870 \pm 0.004 $ & $1.62 \pm 0.04 $
		\end{tabular}
	\end{ruledtabular}
	
\end{table}

\section{Conclusion \label{sec:conclusion}}

In conclusion, we have experimentally implemented an adaptive procedure for quantum process tomography of single qubit states. This procedure fits in the framework of Bayesian quantum tomography and is based on self-learning measurements, which are chosen according to the criterion of maximal information gain. Our numerical and experimental results show, that adaptive tomography outperforms the strategy based on random measurements for unitary and close-to-unitary processes. The advantage in performance is qualitative, i.e. the Bures distance of the estimated $\chi$-matrix to the true one scales better with the number of measurements performed. The procedure may be extended to enable the reconstruction of trace-non-preserving processes. In this case we have proposed an approximation for the exact expression for the information gain, allowing a significant computational speed-up, while preserving the advantage of adaptivity. The results allow us to conclude, that adaptive tomography is advantageous for all rank-1 quantum processes, independently of the amount of loss.

We have studied the behavior of tomography under the influence of instrumental noise. By monitoring the chi-squared test statistic we were able to identify the ultimate noise floor even when the true process is unknown. An important observation is that adaptive tomography has lower noise floor than non-adaptive one for the same level of instrumental noise. We believe, that the self-learning strategy may be further tailored to avoid especially noisy measurements and reduce the noise floor even further. This is to be verified in the future works.

Other directions for further research may include incorporating self-learning measurements into more sophisticated process tomography protocols, like ancilla-assisted process tomography, or direct characterization of quantum dynamics \cite{Lidar_PRL06}. One may also envisage the application of active learning techniques for suppression of technical noise and source drifts, inevitable in any quantum experiment.  

\begin{acknowledgments}
	This work was supported by the Russian Science Foundation (project 16-12-00017).
\end{acknowledgments}

\appendix

\section{Sampling}
Bayesian inference requires calculation of high-dimensional integrals in the relations (\ref{BayesAve}), (\ref{BayesDistrSize}) or when normalizing the posterior distribution. This is a computationally extensive task. In order to circumvent this difficulty we use an approximation technique based on Markov chain Monte Carlo (MCMC) methods~\cite{Doucet_01}. In \emph{sequential importance sampling} (SIS) the posterior distribution $p(\chi|\mathcal{D})$ is replaced by a set of samples $\{\chi_s\}$. Each sample has its weight $w_s$, and the posterior distribution is approximated as follows:
\begin{equation}
p(\chi|\mathcal{D})= \sum\limits_{s=1}^{S} w_s \delta(\chi-\chi_s),
\end{equation}
where $S$ is a total number of samples. Sample positions are fixed and only weights are updated with the data received according to the following recurrent rule~\cite{Houlsby_PRA12}:
\begin{equation}
w_s^{(n+1)} \propto w_s^{(n)}\mathbb{P}(\gamma_{n+1}|\chi_s,\alpha_{n+1}),
\end{equation}
where a proportional multiplier is chosen to satisfy the normalization constraint $\sum_{s=1}^{S} w_s = 1$. Such a fast numerical procedure allows an adaptive strategy to operate and control the experimental apparatus in real time. We use $10^3$ samples (or $10^4$ for trace non-preserving processes) for the $d = 2$ case. The higher is the dimension of the space, the more samples should be used for a good approximation.

\paragraph*{$\chi$-matrix generation.} SIS requires an efficient way to generate random samples~$\{\chi_s\}$. Let us first consider trace-preserving processes. Our generation method is based on the fact that operation elements~$\{E_k\}$ constitute a unitary block matrix~$U$~(\ref{BlockMatrixForm}). Provided a unitary matrix~$U$, operation elements~$\{E_k\}$ are found and then a $\chi$-matrix is obtained. A uniformly distributed (i.e. Haar) random unitary matrix~$U$ can be obtained via a Gram-Schmidt orthogonalization or QR-decomposition of a matrix~$G$, pertaining to the Ginibre ensemble~\cite{Mezzadri_AMS07}: $G = U R$ (here $R$ is a matrix, irrelevant to our discussion). $G$ by definition has independent and identically distributed (i.\,i.\,d.) random Gaussian matrix elements with zero mean. In practice it is sufficient to generate~$U$ of $d^3 \times d$ size (and $G$ accordingly), because only the first ``block column'' of $U$ determines~$\chi$.

Generation of trace non-preserving $\chi$-matrices is slightly different. Again, we exploit the relation~(\ref{BlockMatrixForm}), but together with the following fact.  Suppose a trace non-preserving process~$\mathcal E_\text{n-p}$ has operator elements~$E_1, \dots, E_{d^2}$, obeying $\sum_{k=1}^{d^2} E_k^\dagger E_k= Q < I$. One can always append an auxiliary operator element $E_{d^2+1}$ to obtain a trace-preserving process~$\mathcal E_\text{p}$. Indeed, $\sum_{k=1}^{d^2} E_k^\dagger E_k + \Delta Q = I$, where $\Delta Q = I - Q$ is a positive semidefinite operator. Performing a Cholesky-like decomposition $\Delta Q = E_{d^2+1}^\dagger E_{d^2+1}$, we find the auxiliary element~$E_{d^2+1}$. Consequently, a random unitary matrix~$U$ of $(d^3+d) \times d$ size is generated. After all corresponding $d^2+1$ operator elements are retrieved from~$U$, the last one is neglected, and a trace non-preserving $\chi$-matrix is generated.

\paragraph*{Resampling.} Unfortunately, while the distribution converges to the true process matrix, the approximation becomes less and less efficient because more samples are assigned with negligible weights. One should monitor an \emph{effective number of samples} $S_\text{eff}=\bigl( \sum_{s=1}^{S}w_s^2 \bigr)^{-1}$. This value can be increased by redistributing the samples. When $S_\text{eff} < 0.1 S$ we perform a resampling procedure which consists of the following steps:
\begin{enumerate}
	\item Include the sample $\chi_s$ to the new set of samples with the probability equal to its weight $w_s$. Stop when the new set of samples has the size of $S$;
	\item Assign equal values to the new weights $w_s:=1/S$;
	\item Perform a random walk for each new sample according to the Metropolis-Hastings algorithm \cite{Hastings_Bio70} to make a correct approximation of the distribution $p(\chi|\mathcal{D})$.  
\end{enumerate}
The third step requires a full likelihood function calculation~(\ref{Posterior}) and a random step procedure generating a valid $\chi$-matrix $\chi'$ in the vicinity of the old one $\chi$. The random step procedure is closely related to the $\chi$-matrix generation process described above. The idea is to retrieve the unitary matrix~$U$~(\ref{BlockMatrixForm}) from a given $\chi$-matrix, then add a random deviation~$dU$ to this matrix, $U' = U + dU$, make $U'$ unitary again via a QR-decomposition, and finally calculate the new sample $\chi'$ corresponding to $U'$. If~$dU$ is ``small'', then from continuity considerations~$\chi'$ lies in the vicinity of the old matrix~$\chi$.

In our implementation~$dU$ belongs to the Ginibre ensemble. The standard deviation of its matrix elements depends on the distribution size to ensure an approximately constant acceptance ratio in the Metropolis-Hastings routine. The unitary matrix~$U$, corresponding to the $\chi$-matrix, can be found by the eigenvalue decomposition of~$\chi$~\cite{NielsenChuang}: $\chi_{mn} = \sum_{ik} V_{mi} \delta_{ik} \lambda_i V_{nk}^*$. Substituting this into~(\ref{ChiMatrixForm}) and comparing with~(\ref{KrausForm}), one can conclude that the operator elements satisfy $E_k = \sqrt \lambda_k \sum_m V_{mk} \tilde E_m$. These operator elements are used to compose the unitary~$U$. For trace non-preserving processes we have to store the last auxiliary operator element~$E_{d^2+1}$, generated during the initialization, separately.

The procedure described above seems to be awkward, and we believe there is scope for its optimization. Moreover, the question about the $\chi$-matrix distribution, which this method induces, remains open. One of the good alternatives is to adopt methods from QETLAB~\cite{qetlab}.

\section{Distribution size as a figure of merit}

\begin{figure}
	\centering
	\includegraphics[width=1.0\linewidth]{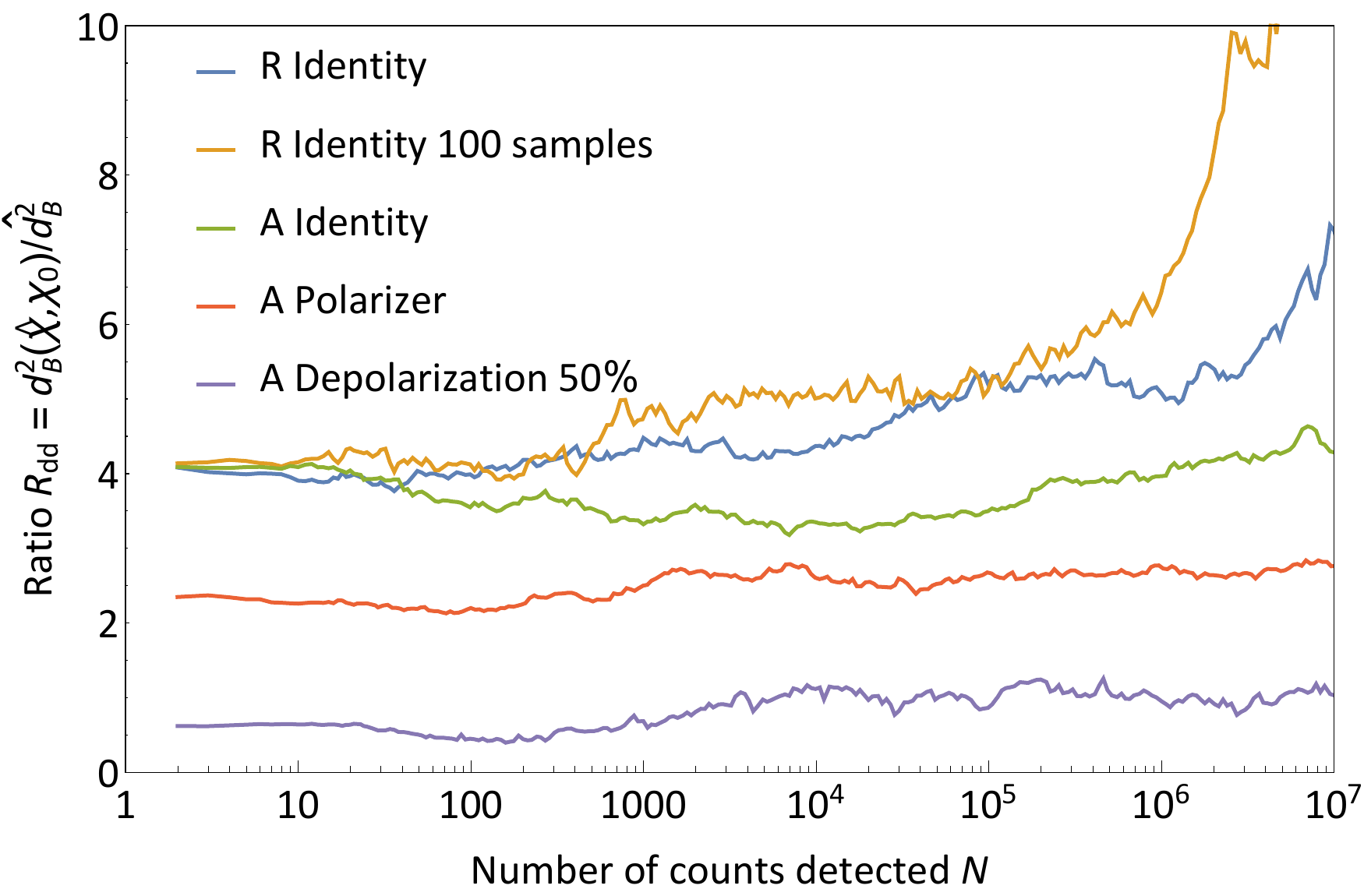}
	\caption{Evolution of the ratio of the distance to the true process to the distribution size $R_{dd}=d_B^2(\hat{\chi},\chi_0)/\overline{d}^2_B$ with the number of registered events $N$ for different true processes. 1000 samples in the approximating distribution are used for every curve except ``R Identity 100 samples'' where 100 samples are used. The results are averaged over 10 runs.}
	\label{fig:sim_rdd_1}
\end{figure}

There is no information about the true process in the experiment, so we suggest to monitor the distribution size $\overline{d}^2_B$ (\ref{BayesDistrSize}) to judge about the convergence of tomography. As shown in Fig.~\ref{fig:sim_rdd_1}, the ratio of the distance to the true process to the distribution size $R_{dd}=d_B^2(\hat{\chi},\chi_0)/\overline{d}^2_B$ keeps a constant value in the course of tomography. This constant value varies depending on the true process, e.g. $R_{dd}\approx 4$ for the identity process. Consequently, one knows that in the experiment the true process in not farther than $R_{dd} \times \overline{d}^2_B$ from the estimator. So if the distribution size $\overline{d}^2_B$ approaches zero, the tomography converges to the true state. However, a sharp growth can be seen for the ``R Identity 100 samples'' curve in Fig.~\ref{fig:sim_rdd_1}. That can be explained by a small amount of samples (100) chosen for the approximation. We use $10^3$ or $10^4$ samples for our simulations and experimental tomography, which is enough for up to $10^6$ registered outcomes.

\section{Recovered process matrices}
\begin{figure*}
\centering
\includegraphics[width=1.0\linewidth]{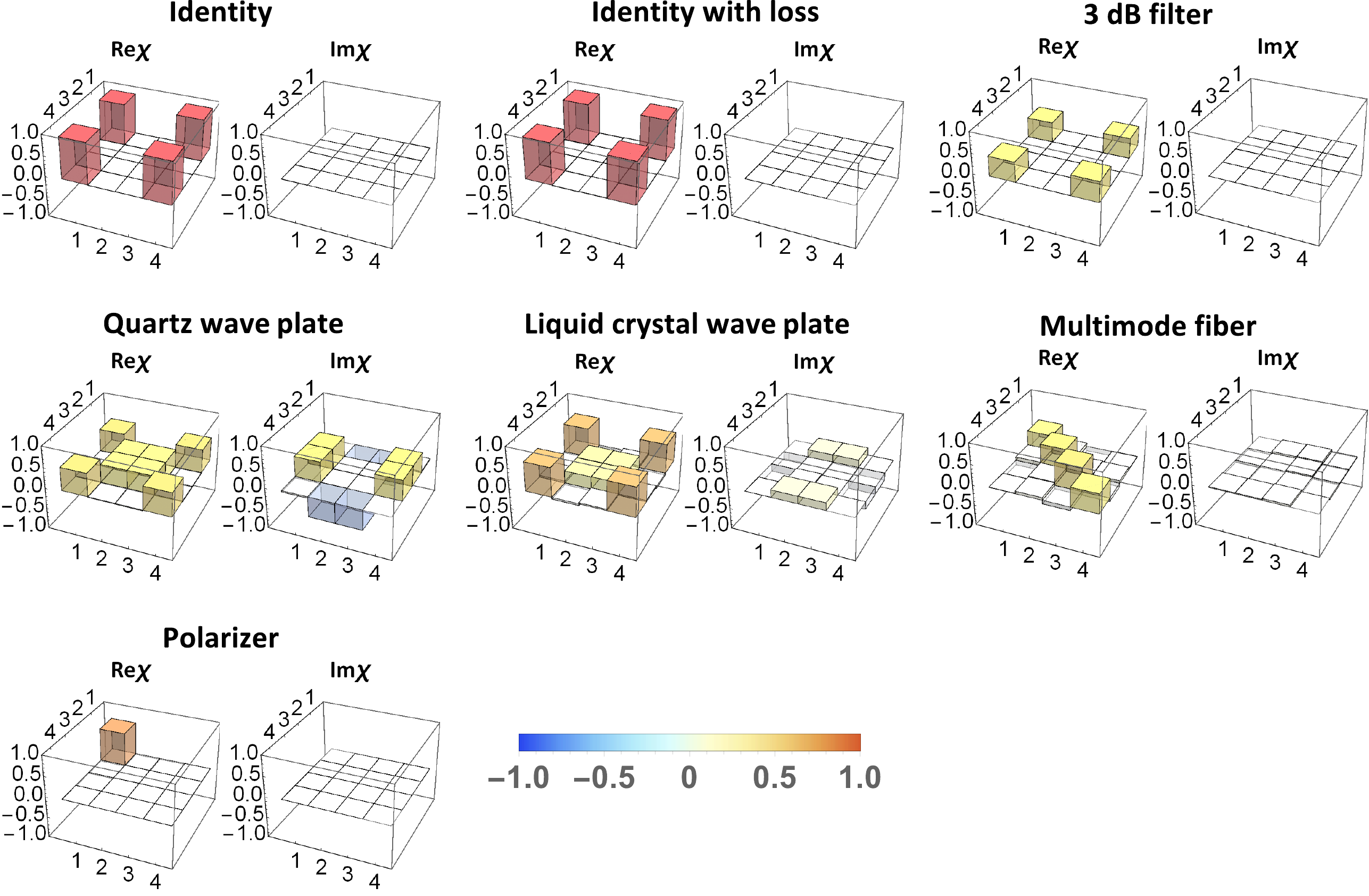}
\caption{Real and imaginary parts of the $\chi$-matrices for different processes recovered via adaptive tomography. Legend explanation: ``Identity with loss'' is the identity process recovered as a trace non-preserving one; ``3\,dB filter'' is an example of a process with polarization insensitive loss, while ``Polarizer'' is a process with strongly polarization-dependent loss; ``Liquid crystal wave plate'' is a partially depolarizing process and ``Multimode fiber'' is an almost completely depolarizing process.}
\label{fig:all_final_states_adaptive}
\end{figure*}

In this section we want to present the $\chi$-matrices for different processes studied experimentally and reconstructed via adaptive tomography. The results are averaged over 10 tomography runs. The processes 1-4 were recovered as trace-preserving ones (see Sec.~\ref{trace-preserving}), on the contrary, processes 5-7 were supposed to be trace non-preserving (see Sec.~\ref{lossy}). All the final $\chi$-matrices are depicted in Fig.~\ref{fig:all_final_states_adaptive}. Some numerical values for the final $\chi$-matrices are given in the following tables for each process. Technical details of the experimental realization of some processes are also discussed here.

\subsection{Identity process}
An ideal identity process has the following $\chi$-matrix:
\begin{equation}
\chi_\text{theor}=\begin{pmatrix}
~1~ & ~0~ & ~0~ & ~1~ \\
0 & 0 & 0 & 0 \\
0 & 0 & 0 & 0 \\
1 & 0 & 0 & 1 \\
\end{pmatrix}. \label{Identity}
\end{equation}
We consider the Bures distance between the ideal and the reconstructed $\chi$-matrices $d_B^2(\hat{\chi},\chi_\text{theor})$, as well as the fidelity between the corresponding Choi-Jamio\l kowski states $F(\hat{\rho},\rho_\text{theor})$.

\begin{table}[h!]
	\begin{ruledtabular}
		\begin{tabular}{lccc}
			\textrm{}&
			\textrm{Value}&
			\textrm{Expected value}\\
			\hline	
			Purity & $0.9907 \pm 0.0008 $ & $1$\rule{0pt}{11pt}\\ 	
			$d_B^2(\hat{\chi},\chi_\text{theor})$ & $0.0094 \pm 0.0008$ & $0$\\ 
			$F(\hat{\rho},\rho_\text{theor})$  & $0.9953 \pm 0.0004$ & $1$ \\
		\end{tabular}
	\end{ruledtabular}
\end{table}

\subsection{Quartz wave plate}
Having the wave-plate $\chi$-matrix at hand, one can recover the angle between the WP axis and the horizontal direction, as well as the phase shift between the orthogonal polarizations. This can be realized by numerical minimization of the Bures distance $d_B(\hat{\chi},\chi_\text{WP}(\phi,\delta))$ between the reconstructed matrix $\hat{\chi}$ and the theoretical wave-plate matrix $\chi_\text{WP}(\phi,\delta)$ using the angle $\phi$ and the phase shift $\delta$ as minimization parameters. The wave-plate chi-matrix $\chi_\text{WP}(\phi,\delta)$ can be obtained using Jones matrix as the first and only operation element $E_1$. The results of the numerical minimization are averaged over 10 tomography runs.

\begin{table}[h]
	\begin{ruledtabular}
		\begin{tabular}{lccc}
			\textrm{}&
			\textrm{Value}&
			\textrm{Expected value}\\
			\hline		
			Angle & $48.4^\circ \pm 1.4 ^\circ $ & $\approx 45^\circ $ \rule{0pt}{11pt}\\ 
			Phase shift & $1.452 \pm 0.016 $ & $ \approx 1.5$ \\
			Purity & $0.9903 \pm 0.0017$ & $1 $\\
		\end{tabular}
	\end{ruledtabular}
\end{table}

\subsection{Multimode fiber\label{sec:MMF}}

A multimode fiber (MMF) was used to experimentally implement a depolarizing channel. The propagating modes in the MMF acquire different phase shifts due to different propagation constants. Due to significant mode-mixing the output light is redistributed between multiple spatial modes, and each spatial mode has its own polarization state. When the spatial mode structure of the beam is averaged by a bucket detector, the measured polarization state is effectively a mixture of the polarization states in each of the modes. So, one can obtain various depolarization degrees depending on fiber length and mode-mixing constants. The MMF we used was long enough and bent strongly enough to provide significant depolarization.

\begin{table}[h]
	\begin{ruledtabular}
		\begin{tabular}{lccc}
			\textrm{}&
			\textrm{Value}\\
			\hline		
			Purity & $0.2754 \pm 0.0024$\rule{0pt}{11pt}\\
		\end{tabular}
	\end{ruledtabular}
	
\end{table}

\subsection{Liquid crystal wave plate}

Implementation of a partially depolarizing channel required another approach. We used a liquid crystal retarder (Thorlabs LCC1111T-B). A variable phase shift of this wave plate can be controlled by a voltage applied. If one keeps the voltage constant the phase shift stays constant too, so the liquid crystal wave plate (LCWP) acts just as a bulk wave plate. The constant voltage $V_0$ can be modified by adding some time-dependent value, e.g. $\Delta V \sin \omega t$, where $\Delta V \ll V_0$. Applying the voltage $V = V_0 + \Delta V \sin \omega t$ one makes LCWP to act as wave plates with slightly different phase shift at different moments of time. If $\omega^{-1}$ is small, compared to the measurement time, the tomography averages all these phase shifts (similar to the spatial average in the previous section), so effectively one can implement a partially-depolarizing channel. The advantage of this approach over using MMF is the fact that a depolarization degree can be easily controlled by the amplitude $\Delta V$.

\begin{table}[h]
	\begin{ruledtabular}
		\begin{tabular}{lccc}
			\textrm{}&
			\textrm{Value}\\
			\hline		
			Purity & $0.648 \pm 0.005$\rule{0pt}{11pt}\\
		\end{tabular}
	\end{ruledtabular}
	
\end{table}

\subsection{Identity process with loss}
Here the identity process was considered as a trace non-preserving one, so the recovered matrix has some loss despite the fact that the theoretical identity process (\ref{Identity}) is lossless.
\begin{table}[h!]
	\begin{ruledtabular}
		\begin{tabular}{lccc}
			\textrm{}&
			\textrm{Value}&
			\textrm{Expected value}\\
			\hline	
			Purity & $0.9854 \pm 0.0025 $ & $1$\rule{0pt}{11pt}\\ 	
			$d_B^2(\hat{\chi},\chi_\text{theor})$ & $0.0148 \pm 0.0026$ & $0$\\ 
			Loss  & $0.060 \pm 0.016$ & $0$ \\
		\end{tabular}
	\end{ruledtabular}
	
\end{table}

\subsection{Neutral filter}

A 3\,dB neutral filter is expected to transmit $\approx 50 \%$ of the light intensity, so we consider the theoretical identity process $\chi$-matrix (\ref{Identity}) multiplied by $0.5$ as a theoretical $\chi$-matrix for the neutral filter: 
\begin{equation}
\chi_\text{theor}=\begin{pmatrix}
~0.5~ & ~0~ & ~0~ & ~0.5~ \\
0 & 0 & 0 & 0 \\
0 & 0 & 0 & 0 \\
0.5 & 0 & 0 & 0.5 \\
\end{pmatrix}.
\end{equation}

\begin{table}[h!]
	\begin{ruledtabular}
		\begin{tabular}{lccc}
			\textrm{}&
			\textrm{Value}&
			\textrm{Expected value}\\
			\hline	
			$d_B^2(\hat{\chi},\chi_\text{theor})$ & $0.0080 \pm 0.0014$ & $0$\rule{0pt}{11pt}\\ 
			Loss  & $51.44 \pm 0.23 \%$ & $50.12 \%$ \\
		\end{tabular}
	\end{ruledtabular}
\end{table}

\subsection{Polarizer}
A polarizer (a nanoparticle linear film polarizer from Thorlabs) was set to transmit only horizontally polarized light, so the only expected non-zero element of the $\chi$-matrix was $\chi_{11}$. Its magnitude corresponds to the transmittance of the horizontal polarization and it was found to be $\approx 77.3\%$ if measured directly. The value for the overall average loss takes into account that vertically polarized light is not transmitted at all.

\begin{table}[h!]
	\begin{ruledtabular}
		\begin{tabular}{lccc}
			\textrm{}&
			\textrm{Value}&
			\textrm{Expected value}\\
			\hline	
			$\hat{\chi}_{11}$ & $0.794 \pm 0.006$ & $\approx 0.773$\rule{0pt}{11pt}\\ 
			Loss  & $60.12 \pm 0.27 \%$ & $\approx 61.35 \%$
		\end{tabular}
	\end{ruledtabular}
\end{table}

\bibliographystyle{apsrev4-1}
\bibliography{ref_base}

\begin{thebibliography}{34}%
\makeatletter
\providecommand \@ifxundefined [1]{%
 \@ifx{#1\undefined}
}%
\providecommand \@ifnum [1]{%
 \ifnum #1\expandafter \@firstoftwo
 \else \expandafter \@secondoftwo
 \fi
}%
\providecommand \@ifx [1]{%
 \ifx #1\expandafter \@firstoftwo
 \else \expandafter \@secondoftwo
 \fi
}%
\providecommand \natexlab [1]{#1}%
\providecommand \enquote  [1]{``#1''}%
\providecommand \bibnamefont  [1]{#1}%
\providecommand \bibfnamefont [1]{#1}%
\providecommand \citenamefont [1]{#1}%
\providecommand \href@noop [0]{\@secondoftwo}%
\providecommand \href [0]{\begingroup \@sanitize@url \@href}%
\providecommand \@href[1]{\@@startlink{#1}\@@href}%
\providecommand \@@href[1]{\endgroup#1\@@endlink}%
\providecommand \@sanitize@url [0]{\catcode `\\12\catcode `\$12\catcode
  `\&12\catcode `\#12\catcode `\^12\catcode `\_12\catcode `\%12\relax}%
\providecommand \@@startlink[1]{}%
\providecommand \@@endlink[0]{}%
\providecommand \url  [0]{\begingroup\@sanitize@url \@url }%
\providecommand \@url [1]{\endgroup\@href {#1}{\urlprefix }}%
\providecommand \urlprefix  [0]{URL }%
\providecommand \Eprint [0]{\href }%
\providecommand \doibase [0]{http://dx.doi.org/}%
\providecommand \selectlanguage [0]{\@gobble}%
\providecommand \bibinfo  [0]{\@secondoftwo}%
\providecommand \bibfield  [0]{\@secondoftwo}%
\providecommand \translation [1]{[#1]}%
\providecommand \BibitemOpen [0]{}%
\providecommand \bibitemStop [0]{}%
\providecommand \bibitemNoStop [0]{.\EOS\space}%
\providecommand \EOS [0]{\spacefactor3000\relax}%
\providecommand \BibitemShut  [1]{\csname bibitem#1\endcsname}%
\let\auto@bib@innerbib\@empty
\bibitem [{\citenamefont {Chuang}\ and\ \citenamefont
  {Nielsen}(1997)}]{Nielsen_JMO97}%
  \BibitemOpen
  \bibfield  {author} {\bibinfo {author} {\bibfnamefont {I.~L.}\ \bibnamefont
  {Chuang}}\ and\ \bibinfo {author} {\bibfnamefont {M.~A.}\ \bibnamefont
  {Nielsen}},\ }\href {\doibase 10.1080/09500349708231894} {\bibfield
  {journal} {\bibinfo  {journal} {Journal of Modern Optics}\ }\textbf {\bibinfo
  {volume} {44}},\ \bibinfo {pages} {2455} (\bibinfo {year}
  {1997})}\BibitemShut {NoStop}%
\bibitem [{\citenamefont {Poyatos}\ \emph {et~al.}(1997)\citenamefont
  {Poyatos}, \citenamefont {Cirac},\ and\ \citenamefont
  {Zoller}}]{Zoller_PRL97}%
  \BibitemOpen
  \bibfield  {author} {\bibinfo {author} {\bibfnamefont {J.~F.}\ \bibnamefont
  {Poyatos}}, \bibinfo {author} {\bibfnamefont {J.~I.}\ \bibnamefont {Cirac}},
  \ and\ \bibinfo {author} {\bibfnamefont {P.}~\bibnamefont {Zoller}},\ }\href
  {\doibase 10.1103/PhysRevLett.78.390} {\bibfield  {journal} {\bibinfo
  {journal} {Phys. Rev. Lett.}\ }\textbf {\bibinfo {volume} {78}},\ \bibinfo
  {pages} {390} (\bibinfo {year} {1997})}\BibitemShut {NoStop}%
\bibitem [{\citenamefont {Nielsen}\ and\ \citenamefont
  {Chuang}(2001)}]{NielsenChuang}%
  \BibitemOpen
  \bibfield  {author} {\bibinfo {author} {\bibfnamefont {M.~A.}\ \bibnamefont
  {Nielsen}}\ and\ \bibinfo {author} {\bibfnamefont {I.~L.}\ \bibnamefont
  {Chuang}},\ }\href@noop {} {\emph {\bibinfo {title} {Quantum Computation and
  Quantum Information}}}\ (\bibinfo  {publisher} {Cambridge University Press},\
  \bibinfo {address} {Cambridge},\ \bibinfo {year} {2001})\BibitemShut
  {NoStop}%
\bibitem [{\citenamefont {Mohseni}\ \emph {et~al.}(2008)\citenamefont
  {Mohseni}, \citenamefont {Rezakhani},\ and\ \citenamefont
  {Lidar}}]{Lidar_PRA08}%
  \BibitemOpen
  \bibfield  {author} {\bibinfo {author} {\bibfnamefont {M.}~\bibnamefont
  {Mohseni}}, \bibinfo {author} {\bibfnamefont {A.~T.}\ \bibnamefont
  {Rezakhani}}, \ and\ \bibinfo {author} {\bibfnamefont {D.~A.}\ \bibnamefont
  {Lidar}},\ }\href {\doibase 10.1103/PhysRevA.77.032322} {\bibfield  {journal}
  {\bibinfo  {journal} {Phys. Rev. A}\ }\textbf {\bibinfo {volume} {77}},\
  \bibinfo {pages} {032322} (\bibinfo {year} {2008})}\BibitemShut {NoStop}%
\bibitem [{\citenamefont {Kravtsov}\ \emph {et~al.}(2013)\citenamefont
  {Kravtsov}, \citenamefont {Straupe}, \citenamefont {Radchenko}, \citenamefont
  {Houlsby}, \citenamefont {Husz\'ar},\ and\ \citenamefont
  {Kulik}}]{Kravtsov_PRA13}%
  \BibitemOpen
  \bibfield  {author} {\bibinfo {author} {\bibfnamefont {K.~S.}\ \bibnamefont
  {Kravtsov}}, \bibinfo {author} {\bibfnamefont {S.~S.}\ \bibnamefont
  {Straupe}}, \bibinfo {author} {\bibfnamefont {I.~V.}\ \bibnamefont
  {Radchenko}}, \bibinfo {author} {\bibfnamefont {N.~M.~T.}\ \bibnamefont
  {Houlsby}}, \bibinfo {author} {\bibfnamefont {F.}~\bibnamefont {Husz\'ar}}, \
  and\ \bibinfo {author} {\bibfnamefont {S.~P.}\ \bibnamefont {Kulik}},\ }\href
  {\doibase 10.1103/PhysRevA.87.062122} {\bibfield  {journal} {\bibinfo
  {journal} {Phys. Rev. A}\ }\textbf {\bibinfo {volume} {87}},\ \bibinfo
  {pages} {062122} (\bibinfo {year} {2013})}\BibitemShut {NoStop}%
\bibitem [{\citenamefont {Mahler}\ \emph {et~al.}(2013)\citenamefont {Mahler},
  \citenamefont {Rozema}, \citenamefont {Darabi}, \citenamefont {Ferrie},
  \citenamefont {Blume-Kohout},\ and\ \citenamefont
  {Steinberg}}]{Steinberg_PRL13}%
  \BibitemOpen
  \bibfield  {author} {\bibinfo {author} {\bibfnamefont {D.~H.}\ \bibnamefont
  {Mahler}}, \bibinfo {author} {\bibfnamefont {L.~A.}\ \bibnamefont {Rozema}},
  \bibinfo {author} {\bibfnamefont {A.}~\bibnamefont {Darabi}}, \bibinfo
  {author} {\bibfnamefont {C.}~\bibnamefont {Ferrie}}, \bibinfo {author}
  {\bibfnamefont {R.}~\bibnamefont {Blume-Kohout}}, \ and\ \bibinfo {author}
  {\bibfnamefont {A.~M.}\ \bibnamefont {Steinberg}},\ }\href {\doibase
  10.1103/PhysRevLett.111.183601} {\bibfield  {journal} {\bibinfo  {journal}
  {Phys. Rev. Lett.}\ }\textbf {\bibinfo {volume} {111}},\ \bibinfo {pages}
  {183601} (\bibinfo {year} {2013})}\BibitemShut {NoStop}%
\bibitem [{\citenamefont {Struchalin}\ \emph {et~al.}(2016)\citenamefont
  {Struchalin}, \citenamefont {Pogorelov}, \citenamefont {Straupe},
  \citenamefont {Kravtsov}, \citenamefont {Radchenko},\ and\ \citenamefont
  {Kulik}}]{Kulik_PRA16}%
  \BibitemOpen
  \bibfield  {author} {\bibinfo {author} {\bibfnamefont {G.~I.}\ \bibnamefont
  {Struchalin}}, \bibinfo {author} {\bibfnamefont {I.~A.}\ \bibnamefont
  {Pogorelov}}, \bibinfo {author} {\bibfnamefont {S.~S.}\ \bibnamefont
  {Straupe}}, \bibinfo {author} {\bibfnamefont {K.~S.}\ \bibnamefont
  {Kravtsov}}, \bibinfo {author} {\bibfnamefont {I.~V.}\ \bibnamefont
  {Radchenko}}, \ and\ \bibinfo {author} {\bibfnamefont {S.~P.}\ \bibnamefont
  {Kulik}},\ }\href {\doibase 10.1103/PhysRevA.93.012103} {\bibfield  {journal}
  {\bibinfo  {journal} {Phys. Rev. A}\ }\textbf {\bibinfo {volume} {93}},\
  \bibinfo {pages} {012103} (\bibinfo {year} {2016})}\BibitemShut {NoStop}%
\bibitem [{\citenamefont {Chapman}\ \emph {et~al.}(2016)\citenamefont
  {Chapman}, \citenamefont {Ferrie},\ and\ \citenamefont
  {Peruzzo}}]{Ferrie_PRL16}%
  \BibitemOpen
  \bibfield  {author} {\bibinfo {author} {\bibfnamefont {R.~J.}\ \bibnamefont
  {Chapman}}, \bibinfo {author} {\bibfnamefont {C.}~\bibnamefont {Ferrie}}, \
  and\ \bibinfo {author} {\bibfnamefont {A.}~\bibnamefont {Peruzzo}},\ }\href
  {\doibase 10.1103/PhysRevLett.117.040402} {\bibfield  {journal} {\bibinfo
  {journal} {Phys. Rev. Lett.}\ }\textbf {\bibinfo {volume} {117}},\ \bibinfo
  {pages} {040402} (\bibinfo {year} {2016})}\BibitemShut {NoStop}%
\bibitem [{\citenamefont {Hou}\ \emph {et~al.}(2016)\citenamefont {Hou},
  \citenamefont {Zhu}, \citenamefont {Xiang}, \citenamefont {Li},\ and\
  \citenamefont {Guo}}]{Guo_NPJQI16}%
  \BibitemOpen
  \bibfield  {author} {\bibinfo {author} {\bibfnamefont {Z.}~\bibnamefont
  {Hou}}, \bibinfo {author} {\bibfnamefont {H.}~\bibnamefont {Zhu}}, \bibinfo
  {author} {\bibfnamefont {G.-Y.}\ \bibnamefont {Xiang}}, \bibinfo {author}
  {\bibfnamefont {C.-F.}\ \bibnamefont {Li}}, \ and\ \bibinfo {author}
  {\bibfnamefont {G.-C.}\ \bibnamefont {Guo}},\ }\href {\doibase
  10.1038/npjqi.2016.1} {\bibfield  {journal} {\bibinfo  {journal} {npj Quantum
  Information}\ }\textbf {\bibinfo {volume} {2}},\ \bibinfo {pages} {16001}
  (\bibinfo {year} {2016})}\BibitemShut {NoStop}%
\bibitem [{\citenamefont {Fischer}\ \emph {et~al.}(2000)\citenamefont
  {Fischer}, \citenamefont {Kienle},\ and\ \citenamefont
  {Freyberger}}]{Freyberger_PRA00}%
  \BibitemOpen
  \bibfield  {author} {\bibinfo {author} {\bibfnamefont {D.~G.}\ \bibnamefont
  {Fischer}}, \bibinfo {author} {\bibfnamefont {S.~H.}\ \bibnamefont {Kienle}},
  \ and\ \bibinfo {author} {\bibfnamefont {M.}~\bibnamefont {Freyberger}},\
  }\href {\doibase 10.1103/PhysRevA.61.032306} {\bibfield  {journal} {\bibinfo
  {journal} {Phys. Rev. A}\ }\textbf {\bibinfo {volume} {61}},\ \bibinfo
  {pages} {032306} (\bibinfo {year} {2000})}\BibitemShut {NoStop}%
\bibitem [{\citenamefont {Hannemann}\ \emph {et~al.}(2002)\citenamefont
  {Hannemann}, \citenamefont {Reiss}, \citenamefont {Balzer}, \citenamefont
  {Neuhauser}, \citenamefont {Toschek},\ and\ \citenamefont
  {Wunderlich}}]{Wunderlich_PRA02}%
  \BibitemOpen
  \bibfield  {author} {\bibinfo {author} {\bibfnamefont {T.}~\bibnamefont
  {Hannemann}}, \bibinfo {author} {\bibfnamefont {D.}~\bibnamefont {Reiss}},
  \bibinfo {author} {\bibfnamefont {C.}~\bibnamefont {Balzer}}, \bibinfo
  {author} {\bibfnamefont {W.}~\bibnamefont {Neuhauser}}, \bibinfo {author}
  {\bibfnamefont {P.~E.}\ \bibnamefont {Toschek}}, \ and\ \bibinfo {author}
  {\bibfnamefont {C.}~\bibnamefont {Wunderlich}},\ }\href {\doibase
  10.1103/PhysRevA.65.050303} {\bibfield  {journal} {\bibinfo  {journal} {Phys.
  Rev. A}\ }\textbf {\bibinfo {volume} {65}},\ \bibinfo {pages} {050303}
  (\bibinfo {year} {2002})}\BibitemShut {NoStop}%
\bibitem [{\citenamefont {Husz{\'a}r}\ and\ \citenamefont
  {Houlsby}(2012)}]{Houlsby_PRA12}%
  \BibitemOpen
  \bibfield  {author} {\bibinfo {author} {\bibfnamefont {F.}~\bibnamefont
  {Husz{\'a}r}}\ and\ \bibinfo {author} {\bibfnamefont {N.~M.~T.}\ \bibnamefont
  {Houlsby}},\ }\href {\doibase 10.1103/PhysRevA.85.052120} {\bibfield
  {journal} {\bibinfo  {journal} {Phys. Rev. A}\ }\textbf {\bibinfo {volume}
  {85}},\ \bibinfo {pages} {052120} (\bibinfo {year} {2012})}\BibitemShut
  {NoStop}%
\bibitem [{\citenamefont {Granade}\ \emph {et~al.}(2012)\citenamefont
  {Granade}, \citenamefont {Ferrie}, \citenamefont {Wiebe},\ and\ \citenamefont
  {Cory}}]{Granade_NJP12}%
  \BibitemOpen
  \bibfield  {author} {\bibinfo {author} {\bibfnamefont {C.~E.}\ \bibnamefont
  {Granade}}, \bibinfo {author} {\bibfnamefont {C.}~\bibnamefont {Ferrie}},
  \bibinfo {author} {\bibfnamefont {N.}~\bibnamefont {Wiebe}}, \ and\ \bibinfo
  {author} {\bibfnamefont {D.~G.}\ \bibnamefont {Cory}},\ }\href
  {http://stacks.iop.org/1367-2630/14/i=10/a=103013} {\bibfield  {journal}
  {\bibinfo  {journal} {New J. Phys.}\ }\textbf {\bibinfo {volume} {14}},\
  \bibinfo {pages} {103013} (\bibinfo {year} {2012})}\BibitemShut {NoStop}%
\bibitem [{\citenamefont {Ferrie}(2014)}]{Ferrie_PRL14}%
  \BibitemOpen
  \bibfield  {author} {\bibinfo {author} {\bibfnamefont {C.}~\bibnamefont
  {Ferrie}},\ }\href {\doibase 10.1103/PhysRevLett.113.190404} {\bibfield
  {journal} {\bibinfo  {journal} {Phys. Rev. Lett.}\ }\textbf {\bibinfo
  {volume} {113}},\ \bibinfo {pages} {190404} (\bibinfo {year}
  {2014})}\BibitemShut {NoStop}%
\bibitem [{\citenamefont {Kalev}\ and\ \citenamefont {Hen}(2015)}]{Hen_NJP15}%
  \BibitemOpen
  \bibfield  {author} {\bibinfo {author} {\bibfnamefont {A.}~\bibnamefont
  {Kalev}}\ and\ \bibinfo {author} {\bibfnamefont {I.}~\bibnamefont {Hen}},\
  }\href {http://stacks.iop.org/1367-2630/17/i=9/a=093008} {\bibfield
  {journal} {\bibinfo  {journal} {New J. Phys.}\ }\textbf {\bibinfo {volume}
  {17}},\ \bibinfo {pages} {093008} (\bibinfo {year} {2015})}\BibitemShut
  {NoStop}%
\bibitem [{\citenamefont {Choi}(1975)}]{Choi1975}%
  \BibitemOpen
  \bibfield  {author} {\bibinfo {author} {\bibfnamefont {M.-D.}\ \bibnamefont
  {Choi}},\ }\href {\doibase http://dx.doi.org/10.1016/0024-3795(75)90075-0}
  {\bibfield  {journal} {\bibinfo  {journal} {Linear Algebra and its
  Applications}\ }\textbf {\bibinfo {volume} {10}},\ \bibinfo {pages} {285 }
  (\bibinfo {year} {1975})}\BibitemShut {NoStop}%
\bibitem [{\citenamefont {Stinespring}(1955)}]{Stinespring_PAMS55}%
  \BibitemOpen
  \bibfield  {author} {\bibinfo {author} {\bibfnamefont {W.~F.}\ \bibnamefont
  {Stinespring}},\ }\href {\doibase 10.1090/S0002-9939-1955-0069403-4}
  {\bibfield  {journal} {\bibinfo  {journal} {Proc. Amer. Math. Soc.}\ }\textbf
  {\bibinfo {volume} {6}},\ \bibinfo {pages} {211} (\bibinfo {year}
  {1955})}\BibitemShut {NoStop}%
\bibitem [{\citenamefont {Granade}\ \emph {et~al.}(2016)\citenamefont
  {Granade}, \citenamefont {Combes},\ and\ \citenamefont
  {Cory}}]{Granade_NJP16}%
  \BibitemOpen
  \bibfield  {author} {\bibinfo {author} {\bibfnamefont {C.}~\bibnamefont
  {Granade}}, \bibinfo {author} {\bibfnamefont {J.}~\bibnamefont {Combes}}, \
  and\ \bibinfo {author} {\bibfnamefont {D.~G.}\ \bibnamefont {Cory}},\ }\href
  {http://stacks.iop.org/1367-2630/18/i=3/a=033024} {\bibfield  {journal}
  {\bibinfo  {journal} {New Journal of Physics}\ }\textbf {\bibinfo {volume}
  {18}},\ \bibinfo {pages} {033024} (\bibinfo {year} {2016})}\BibitemShut
  {NoStop}%
\bibitem [{\citenamefont {Jamio{\l}kowski}(1972)}]{Jamiokowski1972}%
  \BibitemOpen
  \bibfield  {author} {\bibinfo {author} {\bibfnamefont {A.}~\bibnamefont
  {Jamio{\l}kowski}},\ }\href
  {http://linkinghub.elsevier.com/retrieve/pii/0034487772900110} {\bibfield
  {journal} {\bibinfo  {journal} {Reports on Mathematical Physics}\ }\textbf
  {\bibinfo {volume} {3}},\ \bibinfo {pages} {275} (\bibinfo {year}
  {1972})}\BibitemShut {NoStop}%
\bibitem [{\citenamefont {D'Ariano}\ and\ \citenamefont
  {Lo~Presti}(2001)}]{D'Ariano_PRL01}%
  \BibitemOpen
  \bibfield  {author} {\bibinfo {author} {\bibfnamefont {G.~M.}\ \bibnamefont
  {D'Ariano}}\ and\ \bibinfo {author} {\bibfnamefont {P.}~\bibnamefont
  {Lo~Presti}},\ }\href {\doibase 10.1103/PhysRevLett.86.4195} {\bibfield
  {journal} {\bibinfo  {journal} {Phys. Rev. Lett.}\ }\textbf {\bibinfo
  {volume} {86}},\ \bibinfo {pages} {4195} (\bibinfo {year}
  {2001})}\BibitemShut {NoStop}%
\bibitem [{\citenamefont {Leung}(2003)}]{Leung_JMP03}%
  \BibitemOpen
  \bibfield  {author} {\bibinfo {author} {\bibfnamefont {D.~W.}\ \bibnamefont
  {Leung}},\ }\href {\doibase http://dx.doi.org/10.1063/1.1518554} {\bibfield
  {journal} {\bibinfo  {journal} {Journal of Mathematical Physics}\ }\textbf
  {\bibinfo {volume} {44}},\ \bibinfo {pages} {528} (\bibinfo {year}
  {2003})}\BibitemShut {NoStop}%
\bibitem [{\citenamefont {Altepeter}\ \emph {et~al.}(2003)\citenamefont
  {Altepeter}, \citenamefont {Branning}, \citenamefont {Jeffrey}, \citenamefont
  {Wei}, \citenamefont {Kwiat}, \citenamefont {Thew}, \citenamefont {O'Brien},
  \citenamefont {Nielsen},\ and\ \citenamefont {White}}]{Altepeter2003}%
  \BibitemOpen
  \bibfield  {author} {\bibinfo {author} {\bibfnamefont {J.~B.}\ \bibnamefont
  {Altepeter}}, \bibinfo {author} {\bibfnamefont {D.}~\bibnamefont {Branning}},
  \bibinfo {author} {\bibfnamefont {E.}~\bibnamefont {Jeffrey}}, \bibinfo
  {author} {\bibfnamefont {T.~C.}\ \bibnamefont {Wei}}, \bibinfo {author}
  {\bibfnamefont {P.~G.}\ \bibnamefont {Kwiat}}, \bibinfo {author}
  {\bibfnamefont {R.~T.}\ \bibnamefont {Thew}}, \bibinfo {author}
  {\bibfnamefont {J.~L.}\ \bibnamefont {O'Brien}}, \bibinfo {author}
  {\bibfnamefont {M.~A.}\ \bibnamefont {Nielsen}}, \ and\ \bibinfo {author}
  {\bibfnamefont {A.~G.}\ \bibnamefont {White}},\ }\href {\doibase
  10.1103/PhysRevLett.90.193601} {\bibfield  {journal} {\bibinfo  {journal}
  {Phys. Rev. Lett.}\ }\textbf {\bibinfo {volume} {90}},\ \bibinfo {pages}
  {193601} (\bibinfo {year} {2003})}\BibitemShut {NoStop}%
\bibitem [{\citenamefont {Gilchrist}\ \emph {et~al.}(2005)\citenamefont
  {Gilchrist}, \citenamefont {Langford},\ and\ \citenamefont
  {Nielsen}}]{Gilchrist2005}%
  \BibitemOpen
  \bibfield  {author} {\bibinfo {author} {\bibfnamefont {A.}~\bibnamefont
  {Gilchrist}}, \bibinfo {author} {\bibfnamefont {N.~K.}\ \bibnamefont
  {Langford}}, \ and\ \bibinfo {author} {\bibfnamefont {M.~A.}\ \bibnamefont
  {Nielsen}},\ }\href {http://link.aps.org/doi/10.1103/PhysRevA.71.062310}
  {\bibfield  {journal} {\bibinfo  {journal} {Phys. Rev. A}\ }\textbf {\bibinfo
  {volume} {71}},\ \bibinfo {pages} {062310} (\bibinfo {year}
  {2005})}\BibitemShut {NoStop}%
\bibitem [{\citenamefont {Bengtsson}\ and\ \citenamefont
  {\.Zyczkovsky}(2006)}]{Zyczkovsky_book_2006}%
  \BibitemOpen
  \bibfield  {author} {\bibinfo {author} {\bibfnamefont {I.}~\bibnamefont
  {Bengtsson}}\ and\ \bibinfo {author} {\bibfnamefont {K.}~\bibnamefont
  {\.Zyczkovsky}},\ }\href@noop {} {\emph {\bibinfo {title} {Geometry of
  Quantum States}}}\ (\bibinfo  {publisher} {Cambridge University Press},\
  \bibinfo {address} {Cambridge},\ \bibinfo {year} {2006})\BibitemShut
  {NoStop}%
\bibitem [{Note1()}]{Note1}%
  \BibitemOpen
  \bibinfo {note} {However the set $\protect \{M^\chi _{\alpha \gamma }\protect
  \}$ does not form the decomposition of unity, $\DOTSB \sum@ \slimits@
  _{\gamma } M^\chi _{\alpha \gamma } \not =I$, unlike the case of state
  tomography, where it is usually assumed that $\DOTSB \sum@ \slimits@ _{\gamma
  } M_{\alpha \gamma } = I$}\BibitemShut {NoStop}%
\bibitem [{\citenamefont {Blume-Kohout}(2010)}]{BlumeKohout_NJP10}%
  \BibitemOpen
  \bibfield  {author} {\bibinfo {author} {\bibfnamefont {R.}~\bibnamefont
  {Blume-Kohout}},\ }\href {http://stacks.iop.org/1367-2630/12/i=4/a=043034}
  {\bibfield  {journal} {\bibinfo  {journal} {New J. Phys.}\ }\textbf {\bibinfo
  {volume} {12}},\ \bibinfo {pages} {043034} (\bibinfo {year}
  {2010})}\BibitemShut {NoStop}%
\bibitem [{\citenamefont {Bogdanov}\ \emph {et~al.}(2011)\citenamefont
  {Bogdanov}, \citenamefont {Brida}, \citenamefont {Bukeev}, \citenamefont
  {Genovese}, \citenamefont {Kravtsov}, \citenamefont {Kulik}, \citenamefont
  {Moreva}, \citenamefont {Soloviev},\ and\ \citenamefont
  {Shurupov}}]{Genovese11}%
  \BibitemOpen
  \bibfield  {author} {\bibinfo {author} {\bibfnamefont {Y.~I.}\ \bibnamefont
  {Bogdanov}}, \bibinfo {author} {\bibfnamefont {G.}~\bibnamefont {Brida}},
  \bibinfo {author} {\bibfnamefont {I.~D.}\ \bibnamefont {Bukeev}}, \bibinfo
  {author} {\bibfnamefont {M.}~\bibnamefont {Genovese}}, \bibinfo {author}
  {\bibfnamefont {K.~S.}\ \bibnamefont {Kravtsov}}, \bibinfo {author}
  {\bibfnamefont {S.~P.}\ \bibnamefont {Kulik}}, \bibinfo {author}
  {\bibfnamefont {E.~V.}\ \bibnamefont {Moreva}}, \bibinfo {author}
  {\bibfnamefont {A.~A.}\ \bibnamefont {Soloviev}}, \ and\ \bibinfo {author}
  {\bibfnamefont {A.~P.}\ \bibnamefont {Shurupov}},\ }\href {\doibase
  10.1103/PhysRevA.84.042108} {\bibfield  {journal} {\bibinfo  {journal} {Phys.
  Rev. A}\ }\textbf {\bibinfo {volume} {84}},\ \bibinfo {pages} {042108}
  (\bibinfo {year} {2011})}\BibitemShut {NoStop}%
\bibitem [{\citenamefont {Mogilevtsev}\ \emph {et~al.}(2013)\citenamefont
  {Mogilevtsev}, \citenamefont {Hradil}, \citenamefont {Rehacek},\ and\
  \citenamefont {Shchesnovich}}]{Mogilevtsev_PRL13}%
  \BibitemOpen
  \bibfield  {author} {\bibinfo {author} {\bibfnamefont {D.}~\bibnamefont
  {Mogilevtsev}}, \bibinfo {author} {\bibfnamefont {Z.}~\bibnamefont {Hradil}},
  \bibinfo {author} {\bibfnamefont {J.}~\bibnamefont {Rehacek}}, \ and\
  \bibinfo {author} {\bibfnamefont {V.~S.}\ \bibnamefont {Shchesnovich}},\
  }\href {\doibase 10.1103/PhysRevLett.111.120403} {\bibfield  {journal}
  {\bibinfo  {journal} {Phys. Rev. Lett.}\ }\textbf {\bibinfo {volume} {111}},\
  \bibinfo {pages} {120403} (\bibinfo {year} {2013})}\BibitemShut {NoStop}%
\bibitem [{\citenamefont {Kim}\ \emph {et~al.}(2006)\citenamefont {Kim},
  \citenamefont {Fiorentino},\ and\ \citenamefont {Wong}}]{Kim2006a}%
  \BibitemOpen
  \bibfield  {author} {\bibinfo {author} {\bibfnamefont {T.}~\bibnamefont
  {Kim}}, \bibinfo {author} {\bibfnamefont {M.}~\bibnamefont {Fiorentino}}, \
  and\ \bibinfo {author} {\bibfnamefont {F.~N.~C.}\ \bibnamefont {Wong}},\
  }\href {\doibase 10.1103/PhysRevA.73.012316} {\bibfield  {journal} {\bibinfo
  {journal} {Physical Review A}\ }\textbf {\bibinfo {volume} {73}},\ \bibinfo
  {pages} {012316} (\bibinfo {year} {2006})}\BibitemShut {NoStop}%
\bibitem [{\citenamefont {Mohseni}\ and\ \citenamefont
  {Lidar}(2006)}]{Lidar_PRL06}%
  \BibitemOpen
  \bibfield  {author} {\bibinfo {author} {\bibfnamefont {M.}~\bibnamefont
  {Mohseni}}\ and\ \bibinfo {author} {\bibfnamefont {D.~A.}\ \bibnamefont
  {Lidar}},\ }\href {\doibase 10.1103/PhysRevLett.97.170501} {\bibfield
  {journal} {\bibinfo  {journal} {Phys. Rev. Lett.}\ }\textbf {\bibinfo
  {volume} {97}},\ \bibinfo {pages} {170501} (\bibinfo {year}
  {2006})}\BibitemShut {NoStop}%
\bibitem [{\citenamefont {Doucet}\ \emph {et~al.}(2001)\citenamefont {Doucet},
  \citenamefont {de~Freitas},\ and\ \citenamefont {Gordon}}]{Doucet_01}%
  \BibitemOpen
  \bibfield  {author} {\bibinfo {author} {\bibfnamefont {A.}~\bibnamefont
  {Doucet}}, \bibinfo {author} {\bibfnamefont {N.}~\bibnamefont {de~Freitas}},
  \ and\ \bibinfo {author} {\bibfnamefont {N.}~\bibnamefont {Gordon}},\
  }\href@noop {} {\emph {\bibinfo {title} {Sequential {Monte-Carlo} in
  Practice}}}\ (\bibinfo  {publisher} {Springer-Verlag},\ \bibinfo {year}
  {2001})\BibitemShut {NoStop}%
\bibitem [{\citenamefont {Mezzadri}(2007)}]{Mezzadri_AMS07}%
  \BibitemOpen
  \bibfield  {author} {\bibinfo {author} {\bibfnamefont {F.}~\bibnamefont
  {Mezzadri}},\ }\href {http://www.ams.org/notices/200705/fea-mezzadri-web.pdf}
  {\bibfield  {journal} {\bibinfo  {journal} {Notices Am. Math. Soc.}\ }\textbf
  {\bibinfo {volume} {54}},\ \bibinfo {pages} {592} (\bibinfo {year}
  {2007})}\BibitemShut {NoStop}%
\bibitem [{\citenamefont {Hastings}(1970)}]{Hastings_Bio70}%
  \BibitemOpen
  \bibfield  {author} {\bibinfo {author} {\bibfnamefont {W.~K.}\ \bibnamefont
  {Hastings}},\ }\href {\doibase 10.1093/biomet/57.1.97} {\bibfield  {journal}
  {\bibinfo  {journal} {Biometrika}\ }\textbf {\bibinfo {volume} {57}},\
  \bibinfo {pages} {97} (\bibinfo {year} {1970})}\BibitemShut {NoStop}%
\bibitem [{\citenamefont {Johnston}(2016)}]{qetlab}%
  \BibitemOpen
  \bibfield  {author} {\bibinfo {author} {\bibfnamefont {N.}~\bibnamefont
  {Johnston}},\ }\href {\doibase 10.5281/zenodo.44637} {\enquote {\bibinfo
  {title} {{QETLAB}: A {MATLAB} toolbox for quantum entanglement, version
  0.9},}\ }\bibinfo {howpublished} {\url{http://qetlab.com}} (\bibinfo {year}
  {2016})\BibitemShut {NoStop}%
\end{thebibliography}%

\end{document}